\documentclass[twocolumn]{aastex62}

\received{June 25, 2019}
\revised{December 5, 2019}
\accepted{December 7, 2019}

\submitjournal{ApJ}

\shorttitle{SYSTEMATIC STUDY OF THE PEAK ENERGY OF GRB SPECTRA}
\shortauthors{Katsukura et al.}

\begin{document}
\title{SYSTEMATIC STUDY OF THE PEAK ENERGY OF THE BROAD-BAND GAMMA-RAY BURST SPECTRA}


\author[1234-5678-1234-5678]{DAISUKE KATSUKURA}
\affil{Department of Physics, Saitama University, Shimo-Okubo, Sakura-ku, Saitama, 338-8570, Japan}\email{katsukura@heal.phy.saitama-u.ac.jp}

\author{TAKANORI SAKAMOTO}
\affiliation{Department of Physics and Mathematics, College of Science and Engineering, Aoyama Gakuin University, 5-10-1 Fuchinobe,\\ Chuo-ku, Sagamihara-shi, Kanagawa 252-5258, Japan}

\author{MAKOTO TASHIRO}
\affil{Department of Physics, Saitama University, Shimo-Okubo, Sakura-ku, Saitama, 338-8570, Japan}

\author{YUKIKATSU TERADA}
\affil{Department of Physics, Saitama University, Shimo-Okubo, Sakura-ku, Saitama, 338-8570, Japan}



\begin{abstract}

   We have performed a systematic study of Gamma-Ray Bursts (GRBs), which have various values in the peak energy of the ${\nu}F_{\nu}$ spectrum of the prompt emission, $E_{{\rm peak}}$, observed by \textsl{Swift}/BAT and \textsl{Fermi}/GBM, investigating  their prompt and  X-ray afterglow emissions. We cataloged the long-lasting GRBs observed by the \textsl{Swift} between 2004 December and 2014 February in 3 categories according to the classification by \citet{2008ApJ...679..570S}: X-Ray Flashes (XRFs), X-Ray Rich GRBs (XRRs), and Classical GRBs (C-GRBs).  We then derived $E^{{\rm obs}}_{{\rm peak}}$, as well as  $E^{{\rm src}}_{{\rm peak}}$  if viable, of the \textsl{Swift} spectra of their prompt emission. We also analyzed their X-Ray afterglows  and  found the trend that the GRB events with a lower $E_{{\rm peak}}^{{\rm src}}$, i.e. softer GRBs,  are fainter in the 0.3--10 keV X-ray luminosity and decay  more slowly than  harder GRBs.  The intrinsic  event rates of the XRFs, XRRs, and C-GRBs were calculated, using the \textsl{Swift}/BAT trigger algorithm.  That of either of the XRRs and XRFs  is larger than that of the C-GRBs. If we assume that the observational diversity of $E_{{\rm peak}}$ is explained  with the off-axis model \citep{2002ApJ...571L..31Y,2004ApJ...607L..103Y}, these results  yield the jet half-opening angle of $\Delta\theta\sim 0.3^\circ$,  and the variance of the observing angles $\theta_{{\rm obs}} \lesssim0.6^{\circ}$. This implies that the tiny variance of the observing angles of $\lesssim0.6^{\circ}$ would be responsible for the $E_{{\rm peak}}$ diversity observed by \textsl{Swift}/BAT, which is unrealistic. Therefore, we conclude that the $E_{{\rm peak}}$ diversity is not explained with the off-axis model, but is likely to originate from some intrinsic properties of the jets.
  
  


\end{abstract}


\section{Introduction} \label{sec:intro}

The Gamma-Ray Bursts (GRBs) whose prompt emission lasts over 2 seconds are called a long GRBs. According to the widely accepted model of the long GRB,  when a massive star dies and prompts a supernova, a black hole and ultra-relativistic jets are formed  and then a long GRB may be observed if the jets point to us \citep{1999Natur.401..453B, 2006ARA&A..44..507W}. The parameter $E_{{\rm peak}}$, which is the peak energy of a $\nu F_{\nu}$ spectrum, indicates the general spectral property of GRBs. Past observations  with \textsl{High Energy Transient Explorer 2} (\textsl{HETE-2}) showed that  $E_{{\rm peak}}$ is distributed  over a broad energy band from keV to MeV \citep{2004ApJ...602..875S}.  Notably, the GRBs observed by \textsl{HETE-2} were classified with following 3 categories in the basis of their softness ratio. They are classical GRBs (C-GRBs), X-ray rich GRBs (XRRs) and X-ray flushes \citep[XRFs;][]{2003AIPC..662..229H,  2003A&A...400.1021B, 2005ApJ...629..311S}, in descending order of softness ratio. The three kinds of bursts are thought to be based on a unified jet picture \citep{2005ApJ...620..355L}.
Various theoretical models have been proposed to explain the emission process of XRFs and the mechanism generating  2--3 orders of  diversity in $E_{peak}$,  including, for example, a high redshift GRB model \citep{2003AIPC..662..229H}, dirty fireball model \citep{1999ApJ...513..656D, 2002MNRAS...332..945R}, GRB jets with a small contrast of Lorentz factors \citep{2005A&A...440..809B}, off-axis jet model \citep{2002ApJ...571L..31Y, 2004ApJ...607L..103Y}, and variable opening-angle model \citep{2005ApJ...620..355L}. The  validity of these models  have been discussed  in conjunction with the  observed data by \textsl{CGRO}/BATSE \citep[e.g.,][]{1999ApJS...122..465P, 2006ApJS...116..29K} and \textsl{HETE-2} \citep[e.g.,][]{2003A&A...400.1021B, 2005ApJ...629..311S}.   However, the information of the prompt emissions  in the available data was insufficient to derive a definite conclusion about the emission mechanism of the long GRBs.    

    The theoretical models to explain the diversity of $E_{{\rm peak}}$ are broadly classified into two genres: (1) $E_{{\rm peak}}$ varies intrinsically from XRF to GRB, and (2) it originates mostly in the geometrical effect, while the intrinsic diversity is limited.  One of the more accepted models for the latter is the off-axis model [e.g., \citet{2002ApJ...571L..31Y, 2004ApJ...607L..103Y}].  The off-axis model explains well at least the smaller end of $E_{{\rm peak}}$. It also expects the afterglow light-curve to include a rising part, which originates in a weak and relativistic beaming effect accompanying the deceleration of the jet when the observer sees the jet off-axis.  Thus, any observational relation between $E_{{\rm peak}}$ in the prompt emissions and afterglow light-curves (e.g., X-ray luminosity and temporal decay index), if found, would give a key to constrain the theoretical model.

  In recent years, the \textsl{ Neil Gehrels Swift Observatory}  \citep{2004ApJ...611..1005G} has been observing the early GRB afterglows in multi-wavelengths  from optical to X-ray bands since its launch in 2004.                \citet{2008ApJ...679..570S} conducted the first systematic study with the early \textsl{Swift} data with regard to the above-mentioned point and suggested that the X-ray luminosity (0.3--10 keV) of small $E_{{\rm peak}}$ events is lower than that of higher $E_{{\rm peak}}$ events.  However, the number of samples in their study was very limited.
  
  In this paper, we report the results of our systematic analysis of prompt and afterglow emissions of long GRBs observed by \textsl{Swift} between 2004 December and 2014 February. We handle them in three categories, following the classification criteria of \citet{2008ApJ...679..570S}: XRFs, XRRs and C-GRBs.  In  \S\ref{sec:analysis}, we describe the samples of GRBs observed by \textsl{Swift} Burst Alert Telescope \citep[BAT;][]{2005SSRv...120..143B} and then the analysis methods of the prompt emission by \textsl{Swift}/BAT and the broad-band afterglows observed by \textsl{Swift} X-Ray Telescope \citep[XRT;][]{2005SSRv...120..165B} and optical telescopes on the ground. In \S\ref{sec:sample}, we explain about details of our samples which were used for analysis of the prompt emissions and  afterglows. In \S\ref{sec:prompt} and \S\ref{sec:afterglow}, we show the results of the systematic analysis of the prompt emissions and  afterglows, respectively. In \S\ref{sec:discussion}, we calculate the intrinsic  event rates of the XRFs, XRRs, and C-GRBs, using the simulator of the \textsl{Swift}/BAT flight-trigger algorithm \citep{2014ApJ...783..24L, 2016ApJ...818...55G}, and then discuss  the consistency of theoretical models generating  2--3 orders of  diversity of $E_{{\rm peak}}$ on the basis of the results of prompt emissions, afterglow emissions, and total numbers of the three classes of GRBs in the the whole universe per year, before summarizing our results in \S\ref{sec:conclusion}.  Throughout this paper, the cosmological parameters of $\Omega_m = 0.274$, $\Omega_{\Lambda} = 0.726$, $H_0 = 70.5$ km s$^{-1}$ Mpc$^{-1}$ \citep{2007ApJS...170..377S} are adopted. Error bars are in the 90\% confidence level unless  noted otherwise.

\section{ANALYSIS} \label{sec:analysis}
\subsection{Classifying the GRBs observed by Swift}
We classified the 750 long GRBs observed by  \textsl{Swift} between 2004 December and 2014 February  into three categories with the classification method  by \citet{2008ApJ...679..570S}: XRFs, XRRs and C-GRBs.  The classification uses the ratio of the fluences between 25--50 keV ($S_{25-50{\rm keV}}$) and 50--100 keV ($S_{50-100{\rm keV}}$), as follows.
\begin{eqnarray}\label{eqn:classification_thresh}
S_{25-50{\rm keV}}/S_{50-100{\rm keV}} & \le & 0.72 {\rm \         (C-GRB)} \nonumber \\
0.72 < S_{25-50{\rm keV}}/S_{50-100{\rm keV}} & \le & 1.32 {\rm \         (XRR)}  \\
S_{25-50{\rm keV}}/S_{50-100{\rm keV}} & > & 1.32 {\rm\         (XRF)} \nonumber 
\end{eqnarray}
We derived the fluences  from the best-fit model of the X-ray spectra presented in \citet{2016ApJ...827..7L} (hereafter BAT3 catalog), and used them for classification. Our samples for the spectral analysis are long GRBs  of which $T_{90}$ in the BAT3 catalog  are longer than 2 sec.  Table~\ref{tbl:num_sample} summarizes the number of GRBs and spectral samples  for each class.

\subsection{Spectral analysis of the prompt emissions}
\subsubsection{Swift/BAT data analysis} \label{ss:bat_analysis}
  All the event data observed by \textsl{Swift}/BAT  were retrieved from HEASARC at NASA Goddard Space Flight Center. The standard BAT software (HEADAS 6.15.1) and the latest calibration database (CALDB: 2009-01-30) at the time of analysis were used. First, we generated the time-averaged spectra (PHA)  from the event data during $t_{100}$\footnote{Time interval from 0\% to 100\% of the total burst fluence}, using the \verb|batgrbproduct| pipeline.   
     Then, systematic errors were added to each spectral data with the command \verb|batupdatephakw|. The energy response functions were generated with the command \verb|batdrmgen|. The command performed the calculation for a fixed single incident angle of the source and we achieved the data of the function if \textsl{Swift} was stationary during the $t_{100}$ interval. As for the data for which the spacecraft slewed during the interval, we generated the response function for every 5 sec.  The counts in each spectrum were weighted-averaged according to the photon count of every 5 sec, using the \verb|addrmf| command \citep{2008ApJ...679..570S}. For the sources classified as C-GRBs, we combined the \textsl{Swift}/BAT data and the data observed by the Gamma-ray Burst Monitor \citep[GBM;][]{2009ApJ...702..791M} onboard \textsl{Fermi}, because  $E_{peak}$ of those GRBs  are expected to lie above the energy range of the \textsl{Swift}/BAT (15--150 keV). We also analyzed some XRRs by combining the BAT data with GBM data if $E_{{\rm peak}}$ value of the XRRs were not constrained by following analysis using only the data of Swift/BAT. \newline \indent  
   Finally, we performed  model fitting  of each  spectrum, using \texttt{xspec}. We used 4 models to fit the spectra: a single power-law (PL), a PL with an exponential cutoff (CPL), Band function \citep{1993ApJ...413..281B}, and constrained Band function \citep[C-Band;][]{2004ApJ...602..875S}. Hereafter,  the chi-squares of PL, CPL, and Band function are referred to as $\chi_{{\rm PL}}^2$, $\chi_{{\rm CPL}}^2$, and $\chi_{{\rm Band}}^2$, respectively. The procedure of the spectrum analysis and our criteria to decide the best-fit model  are as follows.   
\begin{enumerate}
\item The spectral data  are fitted by the following models in the order of PL (two free parameters), CPL (three free parameters), and Band function (four free parameters).
\item We choose, as the best-fit model,
\begin{enumerate}
\item PL if $\chi_{{\rm CPL}}^2-\chi_{{\rm PL}}^2\le6$, or  
\item CPL if $\chi_{{\rm CPL}}^2-\chi_{{\rm PL}}^2\ge6$ and $\chi_{{\rm Band}}^2-\chi_{{\rm CPL}}^2\le6$, or
\item Band function if $\chi_{{\rm CPL}}^2-\chi_{{\rm PL}}^2\ge6$ and $\chi_{{\rm Band}}^2-\chi_{{\rm CPL}}^2\le6$.
\end{enumerate}
\item If the spectrum is best-fitted by PL and if its photon index $\Gamma_{{\rm PL}} < -2$, we fit the  spectrum  further with the C-Band model and give a tighter  constraint on the value of $E_{peak}$.
\end{enumerate}

\subsubsection{Fermi/GBM data analysis}
We retrieved time-tagged event data (TTE) of  some of the XRRs and all the C-GRBs, corresponding to our \textsl{Swift}/BAT samples, observed by \textsl{Fermi}/GBM  from HEASARC at NASA Goddard Space Flight Center.  \textsl{Fermi} science tools version V10r0p5 were used for data reduction. \textsl{Fermi}/GBM has 12 NaI detectors and 2 BGO detectors, which are numbered 0--11 for NaI and 0--1 for BGO. We selected two NaI detectors and one BGO detector  with the following criteria.
\begin{enumerate}
\item We choose the NaI detectors with source angles $\le 60^{\circ}$ \citep{2014ApJS..211...12G}.
\item We  make the light curves for those event data, using the \verb|gtbin| command.
\item The background  is estimated  from the fitting result of the pre- and post-burst light curve data with polynomial functions ($\chi^2$  minimization) (see the following paragraph for detail).
\item We generate the background-subtracted light curves and select two NaI detectors  the data of which have the highest signal-to-noise (SN) ratios.
\item If the selected NaI detector is one of 0--5, we use BGO-0 data, or otherwise  BGO-1 data.
\end{enumerate}
 The exposure of the spectrum of the foreground object was chosen to be $t_{100}$ obtained by \textsl{Swift}/BAT. The energy response functions  were taken from the public archive of \textsl{Fermi} Science Data Center.
 The background (Item 3 in the list above)  was estimated  from the result of the model fitting; a pair of the 1000s light-curves before and after the event, i.e., one from 1000 seconds before the BAT trigger and the other  for 1000 seconds after the end of BAT $t_{100}$, generated from the CSPEC data, for each channel were fitted with 1-4th-order polynomial functions, and then  the best-fit model was incorporated into the burst time-intervals.

  We performed joint spectral analysis with the \textsl{Swift}/BAT and \textsl{Fermi}/GBM data to better constrain the spectral parameters for the hard GRBs. The energy ranges used in the spectral analysis  were 8--1000 keV for NaI and 0.3--38 MeV for BGO \citep{2014ApJS..211...12G}.
 In the  simultaneous spectral fitting,  a constant factor to the model for each dataset relative to the \textsl{Swift}/BAT data was introduced to take into account the uncertainty in the cross-instrumental calibration.   The criteria to decide the best-fit model  were the same as in section \ref{ss:bat_analysis}

\subsection{Analysis of afterglows}
\subsubsection{X-Ray afterglow}
The X-ray afterglow samples are limited to the GRBs that have the well-constrained $E_{{\rm peak}}$  in our sample, following the analyses  described in the previous section. We retrieved X-ray afterglow light-curves (in the 0.3--10 keV band) through the UK \textsl{Swift} Science Data  Centre \citep[][http://www.swift.ac.uk]{2009MNRAS...397..1177E}. In some GRBs, X-ray flares \citep{2006ApJ...642..354Z} occurred during the shallow decay phase in X-ray afterglow, which were excluded by eye inspection from our sample. We made the 0.3--10 keV light curves in luminosity ($L_{0.3-10{\rm keV}}$) for the events with known redshifts  in the analyzed samples, using equation (\ref{eq:xray_lumi}):

\begin{equation} \label{eq:xray_lumi}
  L_{0.3-10{\rm keV}}= 4\pi d_L^2(1+z)^{-\Gamma-2}F_{0.3-10{\rm keV}},
\end{equation}
 where $d_L$ and $F_{0.3-10{\rm keV}}$ are the luminosity distance and the energy flux observed by \textsl{Swift}/XRT, respectively, and $\Gamma$ is the photon index of the X-ray afterglow at the late-time phase, which is available in the UK \textsl{Swift} Science Data  Centre. We then performed  model fitting for X-ray energy flux and luminosity light-curves with the models of a simple power-law (PL), a PL with one temporal break (BPL1), a PL with two temporal breaks (BPL2), and a PL with three temporal breaks (BPL3)  in this order until the resultant $\chi^2$ did not show  an improvement greater than 2. Accordingly,  the X-ray luminosity ($L_{0.3-10{\rm keV,200s}}$) and the temporal decay index ($\Gamma_{200s}$) at 200 seconds after the trigger at the GRB rest frame were derived. Some GRBs  were in the steep decay phase  in the $\Gamma_{200s}$ interval, in which case the temporal index  in the period following  the shallow decay phase was extrapolated to the epoch of the 200 seconds and the values of $L_{0.3-10{\rm keV, }200s}$ and $\Gamma_{200s}$ were derived at that  epoch  in the same way was as described in \citet{2016ApJ...826..45R}.

\subsubsection{Optical afterglow}
 To analyze the optical afterglow, we collected the optical data published in Gamma-ray burst Coordinate Network (GCN)  and literature. Table~\ref{tbl:opt_sample} summarizes the references of our samples. The galactic extinctions are corrected  according to \citet{1998ApJ...500..525S}.


\section{Sample of GRBs} \label{sec:sample}
\subsection{Results of classification of the Swift GRBs}\label{ss:GRB_class}
\begin{figure}[h]
\centering
\includegraphics[clip,width=9cm]{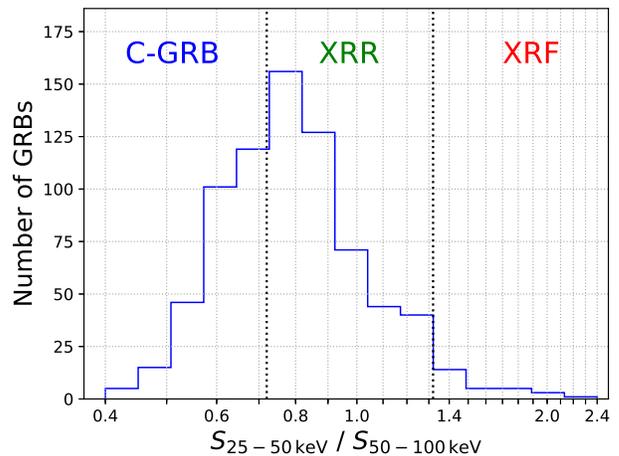}
\figcaption{GRB number histogram for $S_{25-50{\rm keV}}$/$S_{50-100{\rm keV}}$. XRFs, XRRs, and C-GRBs are distributed continuously from the single peak. \label{fig:num_hist}}
\end{figure}

We  cataloged in Table \ref{tbl:num_sample} the long GRBs observed by  \textsl{Swift} between 2004 December and 2014 February  with a classification of XRFs, XRRs, and C-GRBs based on equation \ref{eq:xray_lumi}. Figure \ref{fig:num_hist} shows the distribution of fluence ratio between 25--50 keV and 50--100 keV ($S_{25-50{\rm keV}}$/$S_{50-100{\rm keV}}$). We found that XRFs, XRRs, and C-GRBs are distributed continuously in a single peak.
\newline
\newline

 \begin{deluxetable}{c|ccc}[h]
\tablecaption{Numbers of raw events, those analyzed for spectra and known redshift events for XRFs, XRRs, and C-GRBs \label{tbl:num_sample}}
\tablehead{
  \colhead{Class} & \colhead{Events} & \colhead{Analyzed samples} & \colhead{Redshift samples}
}
\startdata
XRF & $28$ ($3.7\%$) & 26 & 11\\ 
XRR &  $452$ ($60.2\%$) & 41 & 20\\ 
C-GRB & $270$ ($36.0\%$) & 13 & 9\\ \hline
sum & $750$ & 80 & 40\\ 
\enddata
\end{deluxetable}

\subsection{Analyzed samples}\label{ss:GRB_sample}
Because a measurement of the prompt emission parameters (e.g. $E_{{\rm peak}}$) is crucial in this study, we select the samples of the well constrained spectral parameters of the prompt emission.  The samples are also required to have a good quality data of \textsl{Swift}/XRT.  The spectral analysis for XRFs, XRRs, and C-GRBs were performed in the basis of the method described in \ref{ss:bat_analysis}. 

Since the number of samples of XRFs were limited, we selected all XRFs with a good quality data of \textsl{Swift}/XRT.  The XRR samples were selected in descending order of the peak flux (15-150 keV).  In case $E_{{\rm peak}}$ was not constrained by the \textsl{Swift}/BAT data alone, we performed joint spectral analysis of \textsl{Swift}/BAT and \textsl{Fermi}/GBM data if they were commonly detected. We continued the analysis until the samples of XRRs were equivalent to fifty.  Since nine XRRs do not have a good quality of \textsl{Swift}/XRT data, the total number of the analyzed samples for XRRs are 41.  As for C-GRBs, it is expected that those of the $E_{{\rm peak}}$ exceed the upper boundary the BAT energy band of 150 keV \citep{2008ApJ...679..570S}. Therefore, the spectral analysis was performed for the events observed by both \textsl{Swift}/BAT and \textsl{Fermi}/GBM.  Because of the requirement of \textsl{Fermi/GBM} data for the analysis of C-GRBs, the total numbers of the analyzed samples of C-GRBs were significantly reduced to 13.

We also constructed the samples with redshifts from the analyzed samples.  Table \ref{tbl:num_sample} summarized the numbers of our entire samples, analyzed samples and redshift samples for XRFs, XRRs, and C-GRBs.

\section{Results of the prompt emissions} \label{sec:prompt}

\subsection{Spectral analysis}\label{ss:prompt_results}
\begin{figure*}[h]
  \centering
  \includegraphics[clip,width=10cm,angle=270]{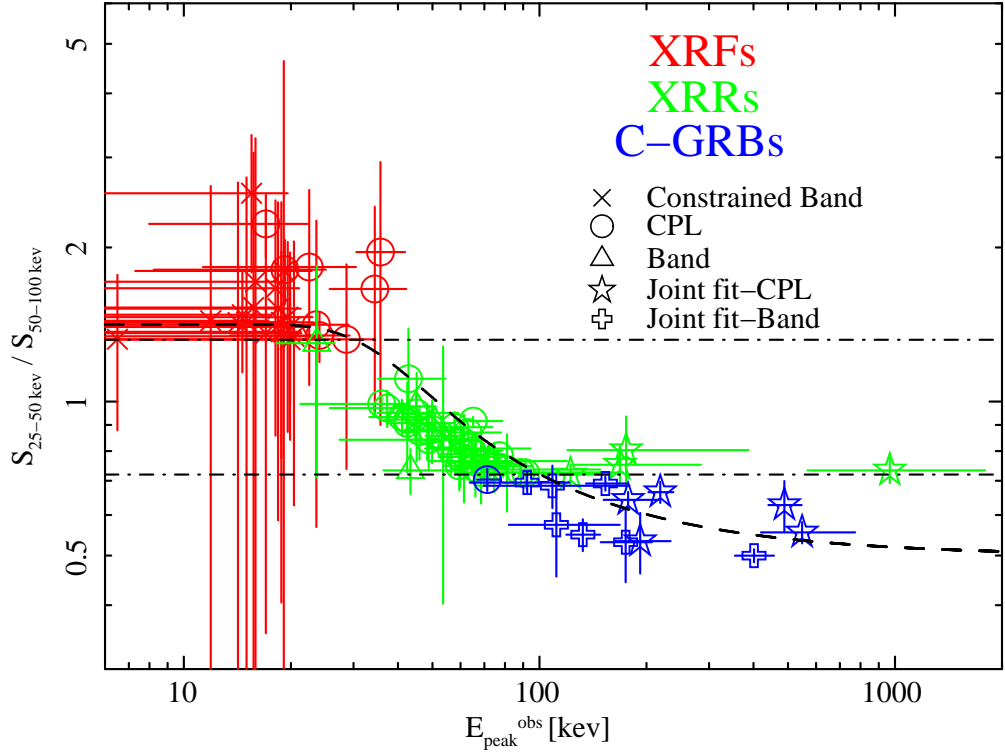}
  \figcaption{Fluence ratio $S_{25-50{\rm keV}}$/$S_{50-100{\rm keV}}$ versus $E_{{\rm peak}}^{{\rm obs}}$. Dashed line  is the theoretical curve of the Band function for  $\alpha=-1$ and the $\beta=-2.5$ (see text). The upper and lower dot-dashed lines  are $S_{25-50{\rm keV}}$/$S_{50-100{\rm keV}}$  for \textbf{$E_{{\rm peak}}^{{\rm obs}}=30$  and  100 keV}, respectively. The number of samples in this figure is $N=80$.\label{fig:ep_ratio}}
\end{figure*}

Table~\ref{tbl:prompt_fit_result} summarizes the  results of the spectral  fitting with the CPL, Band function,  and C-Band  models. Figure \ref{fig:ep_ratio} shows the relation between the fluence ratio $S_{25-50{\rm keV}}$/$S_{50-100{\rm keV}}$ and $E_{{\rm peak}}^{{\rm obs}}$, which is defined as the $E_{{\rm peak}}$ in the observer's frame. The  theoretical curve of $S_{25-50{\rm keV}}$/$S_{50-100{\rm keV}}$  in the case of the low energy spectral index $\alpha=-1$ and the high energy spectral index $\beta=-2.5$  for the Band function is overlaid in the figure (dashed line). The value of $S_{25-50{\rm keV}}$/$S_{50-100{\rm keV}}$ depends on $E_{{\rm peak}}^{{\rm obs}}$ strongly, whereas it is not  a strong function of the values of $\alpha$ and $\beta$.  Therefore, in the energy range of \textsl{Swift}/BAT, this classification according to the fluence ratio is practically equivalent to that  according to $E_{{\rm peak}}^{{\rm obs}}$.  In addition, we note that the $E_{{\rm peak}}^{{\rm obs}}$ of XRFs, XRRs, and C-GRBs are distributed continuously from a few to hundreds of keV.

\begin{figure*}[h]
\centering
\includegraphics[clip,width=10cm,angle=270]{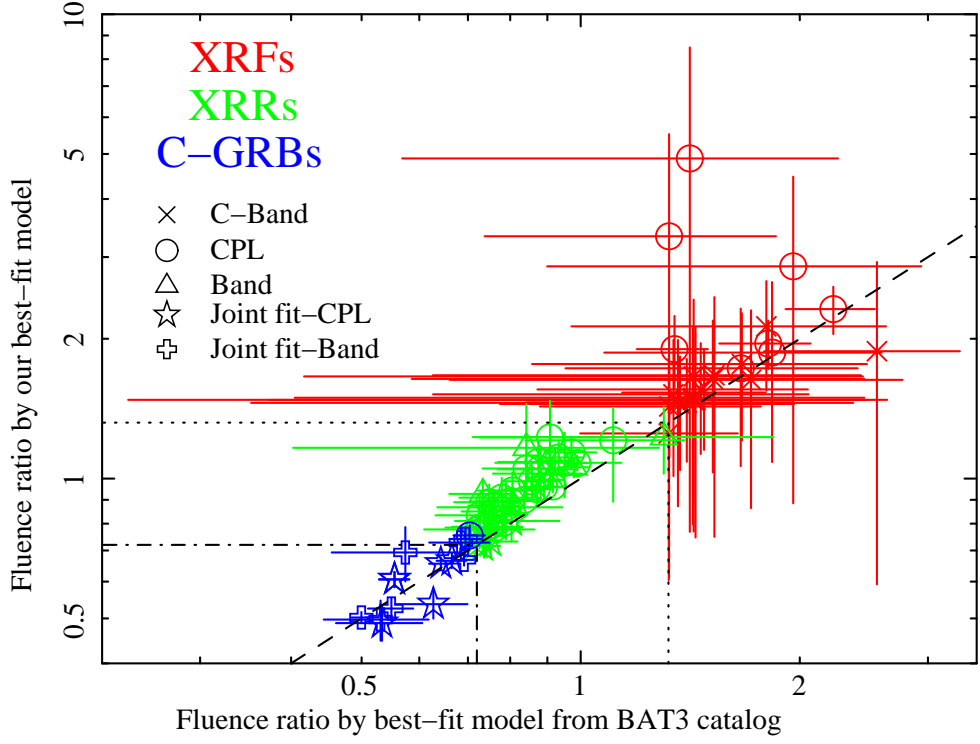}
\figcaption{Comparison of the fluence ratios based on our spectral modelling ($R_{{\rm mod}}$) with those based on the spectral fits of the BAT3 catalog ($R_{{\rm BAT3}}$). The dashed line represents $R_{{\rm mod}} = R_{{\rm BAT3}}$. The dotted and the dot-dashed lines are $R_{{\rm mod}},\,R_{{\rm BAT3}}=1.32$ and $R_{{\rm mod}},\,R_{{\rm BAT3}}=0.72$, respectively. The number of samples in this figure is $N=80$.\label{fig:check_ratio}}
\end{figure*}

Figure \ref{fig:check_ratio} shows comparison of the fluence ratios based on our spectral modelling ($R_{{\rm mod}}$) with those based on the spectral fits of the BAT3 catalog ($R_{{\rm BAT3}}$). $R_{{\rm mod}}$ tends to be slightly larger, especially in XRRs, than $R_{{\rm BAT3}}$. GRB080916A and GRB080714 change to XRR in the basis of classification using $R_{{\rm mod}}$. They are, however, consistent to be C-GRB considering their error regions. Additionally, most of analyzed sample do not deviate significantly from $R_{{\rm mod}}=R_{{\rm BAT3}}$. Thus, the tendency generated by the difference of modeling between BAT3 catalog and our analysis is negligible.

\begin{figure*}[h]
  \centering
  \includegraphics[clip,width=10cm,angle=270]{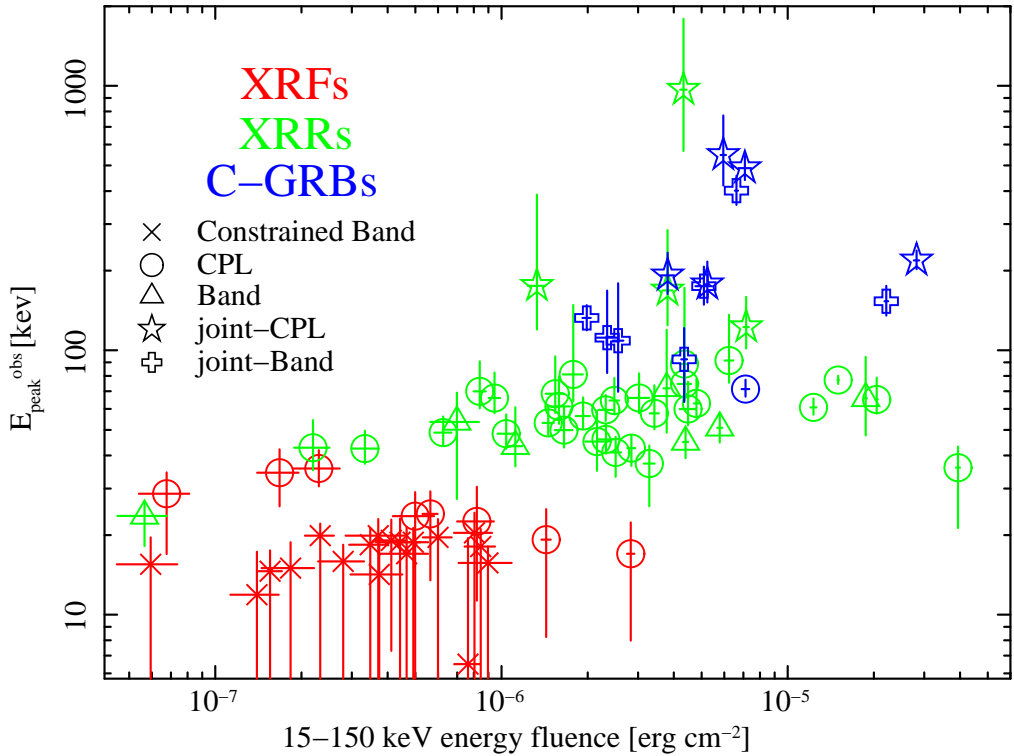}
  \caption{$E_{peak}^{obs}$ versus energy fluence in the energy band of \textsl{Swift}/BAT (15--150 keV). The number of samples in this figure is $N=80$.\label{fig:fuence_ep}}
\end{figure*}

Figure \ref{fig:fuence_ep} shows the energy fluence in the energy band of \textsl{Swift}/BAT (15--150 keV) versus $E_{{\rm peak}}^{{\rm obs}}$. The 15--150 keV fluence of the XRFs tends to be lower (dimmer) than those of the XRRs and C-GRBs.

\begin{figure*}[h]
 \centering
   \includegraphics[clip,width=10cm,angle=270]{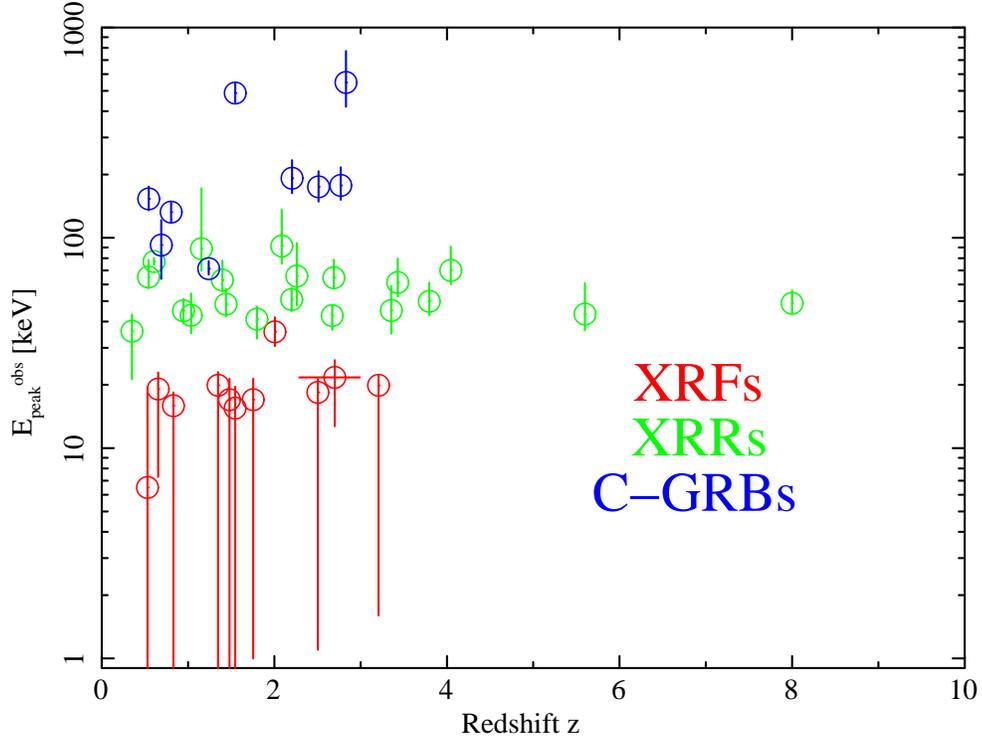}
  \caption{Distribution of $z$ and $E_{peak}^{obs}$. The number of samples in this figure is $N=40$. \label{fig:z_ep_obs}}
\end{figure*}

\begin{figure*}[h]
  \centering
   \includegraphics[clip,width=10cm,angle=270]{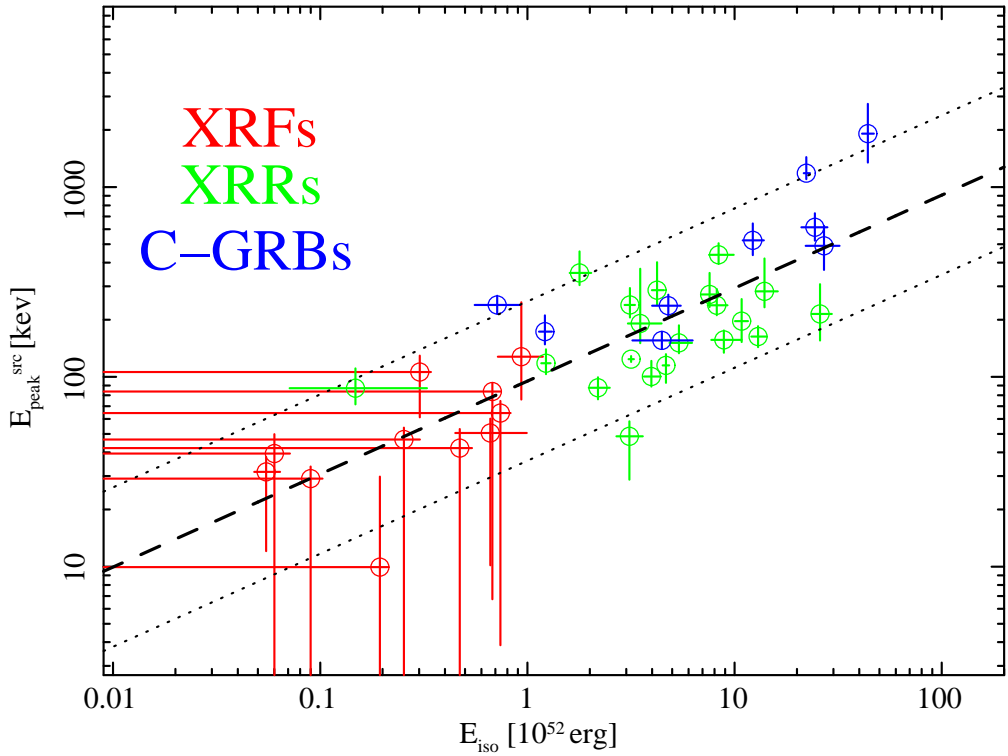}
  \caption{ Scatter plot of $E_{{\rm peak}}^{{\rm src}}$ versus $E_{{\rm iso}}$  for our samples. The dashed line represents the best-fit power-law ($E_{peak}^{{\rm src}}=95\times(E_{{\rm iso}}/10^{52})^{0.49}$) and the two dotted lines  delineate the region corresponding to $\pm 2 \sigma$, which  are derived  from   \citet{2006MNRAS...372..233A}. The number of samples in this figure is $N=40$. \label{fig:amati_relation}}
\end{figure*}

The redshift ($z$), $E_{peak}^{src}$, and the total isotropic-equivalent radiated energy $E_{{\rm iso}}$ of our samples are summarized in table \ref{tbl:intrinsic_prompt_result}. Figure \ref{fig:z_ep_obs} shows the distribution of z and $E_{{\rm peak}}^{{\rm obs}}$. No clear trend  of clustering of the \textsl{Swift}/BAT XRF population, especially towards the high redshift end, is observed, which contradicts the suggestion that the XRFs would be in  high-redshift origin \citep{Heise01}. Figure \ref{fig:amati_relation} shows the  correlation,  known as the Amati relation \citep{2002A&A...390..81A}, between the rest-frame $E_{{\rm peak}}$ ($E_{{\rm peak}}^{{\rm src}}$) and $E_{{\rm iso}}$. Our samples are consistent within the error  with the relation  derived from the best-fit result of \citet{2006MNRAS...372..233A} for the $\pm 2 \sigma$ region except  for XRR130925A.    Note that the figure also shows that our sample has the diversity  for 2 and 3 orders of magnitude in $E_{{\rm peak}}^{{\rm src}}$ and  $E_{{\rm iso}}$, respectively.



\begin{longrotatetable}
\begin{deluxetable}{l c c c c c c c c c c}
\tablewidth{700pt}
\tabletypesize{\small}
\tablecaption{Summary of our spectral fitting results  with the CPL, Band function, and C-Band models.\label{tbl:prompt_fit_result}}
\tablehead{
\colhead{Events} & \colhead{$\alpha$\tablenotemark{a}} & \colhead{$\beta$\tablenotemark{b}} & \colhead{$E_{{\rm peak}}^{{\rm obs}}$} & \colhead{$\chi^2$/d.o.f.} & \colhead{ratio\tablenotemark{c}} & \colhead{fluence\tablenotemark{d}} & \colhead{} & \colhead{constant factor\tablenotemark{e}} & \colhead{} & \colhead{model\tablenotemark{f}} \\
\colhead{} & \colhead{} & \colhead{} & \colhead{(keV)} & \colhead{} &  \colhead{} & \colhead{} &\colhead{GBM-NaI(1)} & \colhead{GBM-NaI(2)} & \colhead{GBM-BGO} & \colhead{} 
}
\startdata
XRF050406 & $> -1.48$ & \nodata & $ 28.7_{-11.7}^{+5.7} $ & 77.42/58 & $ 1.32_{-0.58}^{+0.53} $ & $ 0.778_{-0.017}^{+0.018} $ & \nodata & \nodata & \nodata & CPL \\
XRF050416A & \nodata & $ -3.08_{-0.23}^{+0.21} $ & $ 19.1_{-11.1}^{+3.8} $ & 59.98/58 & $ 2.26_{-0.86}^{+0.76} $ & $ 4.13_{-0.53}^{+0.55} $ & \nodata & \nodata & \nodata & C-Band \\
XRF050819 & \nodata & $ -2.69_{-0.30}^{+0.27} $ & $ 18.4_{-17.1}^{+2.9} $ & 58.49/58 & $ 1.56_{-0.55}^{+0.50} $ & $ 3.48_{-0.53}^{+0.56} $ & \nodata & \nodata & \nodata & C-Band \\
XRF050824 & \nodata & $ -2.72_{-0.39}^{+0.34} $ & $ < 18.4 $ & 55.82/58 & $ 1.73_{-0.82}^{+0.71} $ & $ 2.78_{-0.50}^{+0.53} $ & \nodata & \nodata & \nodata & C-Band \\
XRF060219 & \nodata & $ -2.50_{-0.33}^{+0.29} $ & $ 18.8_{-17.1}^{+4.3} $ & 64.84/58 & $ 1.44_{-0.55}^{+0.51} $ & $ 4.40_{-0.78}^{+0.81} $ & \nodata & \nodata & \nodata & C-Band \\
XRF060428B & \nodata & $ -2.81_{-0.26}^{+0.24} $ & $ < 21.1 $ & 64.77/58 & $ 1.72_{-0.52}^{+0.48} $ & $ 8.441_{-1.03}^{+1.07} $  & \nodata & \nodata & \nodata & C-Band \\
XRF060923B & \nodata & $ -2.53_{-0.25}^{+0.23} $ & $ 18.8_{-17.2}^{+2.5} $ & 55.17/58 & $ 1.39_{-0.37}^{+0.35} $ & $ 4.90_{-0.62}^{+0.63} $ & \nodata & \nodata & \nodata & C-Band \\
XRF060926 & \nodata & $ -2.54_{-0.23}^{+0.21} $ & $ 19.9_{-18.2}^{+2.2} $ & 59.14/58 & $ 1.41_{-0.36}^{+0.34} $ & $ 2.32_{-0.27}^{+0.28} $ & \nodata & \nodata & \nodata & C-Band \\
XRF070330 & $ 0.27_{-1.26}^{+1.79} $ & \nodata & $ 30.4_{-8.7}^{+7.8} $ & 64.72/58 & $ 1.66_{-0.71}^{+0.76} $ & $ 1.67_{-0.30}^{+0.26} $ & \nodata & \nodata & \nodata & CPL \\
XRF070714A & \nodata & $ -2.62_{-0.21}^{+0.22} $ & $ < 17.5 $ & 61.10/58 & $ 1.48_{-0.34}^{+0.32} $ & $ 1.55 \pm 0.16 $ & \nodata & \nodata & \nodata & C-Band \\
XRF080218B & $ > -1.44 $ & \nodata & $ < 29.1 $ & 48.44/58 & $ 1.41 \pm 0.84 $ & $ 4.81_{-0.82}^{+0.94} $ & \nodata & \nodata & \nodata & CPL \\
XRF080520 & \nodata & $ -2.80_{-0.47}^{+0.41} $ & $ 15.5_{-14}^{+4.1} $ & 55.46/58 & $ 2.11_{-1.54}^{+1.22} $ & $ 0.59_{-0.14}^{+0.15} $ & \nodata & \nodata & \nodata & C-Band \\
XRF081007 & \nodata & $ -2.52_{-0.21}^{+0.19} $ & $ < 19.5 $ & 58.28/58 & $ 1.34_{-0.29}^{+0.28} $ & $ 7.61_{-0.81}^{+0.82} $ & \nodata & \nodata & \nodata & C-Band \\
XRF100425A & \nodata & $ -2.48_{-0.32}^{+0.29} $ & $ 17.0_{-16}^{+4.4} $ & 50.20/58 & $ 1.35_{-0.53}^{+0.49} $ & $ 4.64_{-0.86}^{+0.91} $ & \nodata & \nodata & \nodata & C-Band \\
XRF110319A & $ -1.39_{-0.44}^{+0.48} $ & \nodata & $ 19.2_{-10}^{+5.8} $ & 53.33/58 & $ 1.82 \pm 0.20 $ & $ 14.3 \pm 0.7 $ & \nodata & \nodata & \nodata & CPL \\
XRF110808A & \nodata & $ -2.27_{-0.43}^{+0.38} $ & $ 19.9_{-19}^{+3.1} $ & 58.71/58 & $ 1.321 \pm 0.32 $ & $ 3.69_{-0.84}^{+0.88} $ & \nodata & \nodata & \nodata & C-Band \\
XRF111129A & \nodata & $ -2.66_{-0.42}^{+0.36} $ & $ 15.0_{-14}^{+3.8} $ & 52.47/58 & $ 1.44_{-0.66}^{+0.59} $ & $ 1.82_{-0.37}^{+0.40} $ & \nodata & \nodata & \nodata & C-Band \\
XRF120116A & $ -1.36_{-0.43}^{+0.47} $ & \nodata & $ 17.0_{-9.0}^{+5.3} $ & 47.85/58 & $ 2.11 \pm 0.22 $ & $ 28.8 \pm 1.3 $ & \nodata & \nodata & \nodata & CPL \\
XRF120724A & \nodata & $ -2.56_{-0.28}^{+0.25} $ & $ 20.4_{-16}^{+3.9} $ & 52.90/58 & $ 1.35_{-0.41}^{+0.39} $ & $ 8.00_{-1.22}^{+1.25} $ & \nodata & \nodata & \nodata & C-Band \\
XRF120816A & \nodata & $ -2.54_{-0.40}^{+0.34} $ & $ < 19.6 $ & 55.78/58 & $ 1.42_{-0.61}^{+0.57} $ & $ 3.73_{-0.76}^{+0.79} $ & \nodata & \nodata & \nodata & C-Band \\
XRF121108A & $ > -2.15 $ & \nodata & $ 22.5_{-12.2}^{+8.3} $ & 46.58/58 & $ 1.83 \pm 0.76 $ & $ 8.28_{-1.17}^{+1.38} $ & \nodata & \nodata & \nodata & CPL \\
XRF121212A & \nodata & $ -2.56_{-0.42}^{+0.37} $ & $ 11.9_{-11.0}^{+5.4} $ & 57.08/58 & $ 1.42_{-0.63}^{+0.56} $ & $ 1.39_{-0.26}^{+0.28} $ & \nodata & \nodata & \nodata & C-Band \\
XRF130608A & \nodata & $ -2.74_{-0.46}^{+0.39} $ & $ 15.7_{-15.0}^{+2.4} $ & 55.47/58 & $ 1.54_{-0.76}^{+0.67} $ & $ 8.90_{-1.84}^{+1.96} $ & \nodata & \nodata & \nodata & C-Band \\
XRF130612A & $ > -2.0 $ & \nodata & $ 35.7_{-5.2}^{+6.2} $ & 48.28/58 & $ 1.96_{-1.06}^{+0.98} $ & $ 8.90_{-1.84}^{+1.96} $ & \nodata & \nodata & \nodata & CPL \\ 
XRF130812A & $-1.14_{-0.57}^{+0.65}$ & \nodata & $ 24.1_{-10.6}^{+5.2} $ & 63.24/58 & $ 1.35 \pm 0.15 $ & $ 6.27 \pm 0.36 $ & \nodata & \nodata & \nodata & CPL \\
XRF140103A & \nodata & $ -2.64_{-0.24}^{+0.22} $ & $ 19.6_{-16.0}^{+3.4} $ & 72.39/58 & $ 1.51_{-0.38}^{+0.36} $ & $ 6.00_{-0.70}^{+0.71} $ & \nodata & \nodata & \nodata & C-Band \\ \hline
XRR050318 & $ -1.08_{-0.41}^{+0.45} $ & \nodata & $ 48.4_{-6.0}^{+8.9} $ & 54.37/58 & $ 1.04_{-0.11}^{+0.12} $ & $ 10.4_{-0.7}^{+0.8} $ & \nodata & \nodata & \nodata & CPL \\
XRR050410 & $ -0.829_{-0.36}^{+0.39} $ & \nodata & $ 74.6_{-10}^{+20} $ & 60.39/58 & $ 0.779 \pm 0.068 $ & $ 43.5 \pm 2.5 $ & \nodata & \nodata & \nodata & CPL \\
XRR050525A & $ -1.01_{-0.10}^{+0.10} $ & \nodata & $ 77.1_{-2.6}^{+3.0} $ & 20.61/58 & $ 0.792 \pm 0.019 $ & $ 151 \pm 2 $ & \nodata & \nodata & \nodata & CPL \\
XRR050915B & $ -1.39_{-0.29}^{+0.31} $ & \nodata & $ 57.7_{-7.5}^{+15} $ & 57.63/58 & $ 0.948 \pm 0.058 $ & $ 34.0 \pm 1.4 $ & \nodata & \nodata & \nodata & CPL \\
XRR060206 & $ -1.06_{-0.31}^{+0.33} $ & \nodata & $ 70.0_{-9.7}^{+20} $ & 58.63/58 & $ 0.809 \pm 0.061 $ & $ 8.42 \pm 0.44 $ & \nodata & \nodata & \nodata & CPL \\
XRR060707 & $ -0.602_{-0.590}^{+0.680} $ & \nodata & $ 61.3_{-8.7}^{+18} $ & 61.65/58 & $ 0.846_{-0.122}^{+0.123} $ & $ 15.8 \pm 1.5 $ & \nodata & \nodata & \nodata & CPL \\
XRR060825 & $ -1.07_{-0.29}^{+0.32} $ & \nodata & $ 66.0_{-8.2}^{+16} $ & 55.92/58 & $ 0.843_{-0.057}^{+0.058} $ & $ 9.55_{-0.47}^{+0.47} $ & \nodata & \nodata & \nodata & CPL \\
XRR060927 & $ 0.37_{-1.1}^{+1.5} $ & $ -2.01_{-0.30}^{+0.17} $ & $ 43.4_{-7.0}^{+17} $ & 63.93/57 & $ 0.810_{-0.067}^{+0.068} $ & $ 11.2 \pm 0.7 $ & \nodata & \nodata & \nodata & Band \\
XRR061222B & $ -1.22_{-0.53}^{+0.60} $ & \nodata & $ 45.2_{-10}^{+13} $ & 63.24/58 & $ 0.970 \pm 0.129 $ & $ 22.7 \pm 1.8 $ & \nodata & \nodata & \nodata & CPL \\
XRR070612B & $ -0.902_{-0.48}^{+0.54} $ & \nodata & $ 81.0_{-16}^{+67} $ & 39.05/58 & $ 0.736 \pm 0.085 $ & $ 18.2 \pm 1.4 $ & \nodata & \nodata & \nodata & CPL \\
XRR070721A & $ > -0.31 $ & $ -3.33_{-2.81}^{+0.76} $ & $ 23.6_{-5.5}^{+2.9} $ & 49.31/57 & $ 1.27_{-0.59}^{+0.54} $ & $ 0.730_{-0.172}^{+0.182} $ & \nodata & \nodata & \nodata & Band \\
XRR071010B & $ -1.22_{-0.35}^{+0.53} $ & $ < -2.18 $ & $ 45.0_{-5.9}^{+6.1} $ & 31.97/57 & 
$ 0.977 \pm 0.037 $ & $ 46.2 \pm 1.0 $& \nodata & \nodata & \nodata & Band \\
XRR080207 & $ -1.17_{-0.25}^{+0.26} $ & \nodata & $ 91.5_{-16}^{+44} $ & 53.43/58 & $ 0.737 \pm 0.039 $ & $ 64.0 \pm 2.1 $ & \nodata & \nodata & \nodata & CPL \\
XRR080212 & $ -0.274_{-0.58}^{+0.67} $ & \nodata & $ 66.0_{-8.2}^{+15} $ & 51.93/58 & $ 0.741_{-0.089}^{+0.090} $ & $ 30.1 \pm 2.6 $ & \nodata & \nodata & \nodata & CPL \\
XRR080603B & $ -1.16_{-0.27}^{+0.29} $ & \nodata & $ 64.6_{-7.3}^{+13} $ & 61.63/58 & $ 0.849_{-0.051}^{+0.051} $ & $ 24.6 \pm 1.1 $ & \nodata & \nodata & \nodata & CPL \\
XRR081128 & $ -1.03_{-0.42}^{+0.47} $ & \nodata & $ 46.0_{-5.3}^{+6.5} $ & 34.14/58 & $ 1.03 \pm 0.11 $ & $ 23.4 \pm 1.6 $ & \nodata & \nodata & \nodata & CPL \\
XRR081221 & $ -1.18_{-0.30}^{+0.51} $ & \nodata & $ 65.8_{-18}^{+28} $ & 31.03/58 & $ 0.816 \pm 0.021 $ & $ 189 \pm 3 $ & \nodata & \nodata & \nodata & CPL \\
XRR090423 & $ -0.803_{-0.46}^{+0.52} $ & \nodata & $ 48.8_{-5.2}^{+7.4} $ & 41.19/58 & $ 0.980_{-0.108}^{+0.110} $ & $ 6.24_{-0.45}^{+0.46} $ & \nodata & \nodata & \nodata & CPL \\
XRR090429B & $ -0.587_{-0.68}^{+0.82} $ & \nodata & $ 42.5_{-5.3}^{+7.1} $ & 31.04/58 & $ 1.15_{-0.20}^{+0.22} $ & $ 3.29_{-0.34}^{+0.36} $ & \nodata & \nodata & \nodata & CPL \\
XRR090531A & $ -0.874_{-0.41}^{+0.45} $ & \nodata & $ 68.5_{-10}^{+26} $ & 41.64/58 & $ 0.785_{-0.081}^{+0.082} $ & $ 15.4 \pm 1.2 $ & \nodata & \nodata & \nodata & CPL \\
XRR090813 & $ -1.57_{-0.14}^{+0.12} $ & \nodata & $ 175_{-55}^{+212} $ & 90.70/86 & $ 0.812_{-0.091}^{+0.091} $ & $ 13.3 \pm 1.0 $ & $ 1.22_{-0.10}^{+0.11}$  & $ 1.31_{-0.12}^{+0.13} $  & $ < 6.58 $ & joint-CPL \\
XRR090912 & $ -0.936_{-0.42}^{+0.46} $ & \nodata & $ 59.9_{-7.9}^{+16} $ & 38.31/58 & $ 0.841_{-0.083}^{+0.084} $ & $ 44.8 \pm 3.1 $ & \nodata & \nodata & \nodata & CPL \\
XRR100615A & $ -1.56_{-0.19}^{+0.20} $ & \nodata & $ 63.0_{-7.4}^{+14} $ & 35.55/58 & $ 0.910_{-0.034}^{+0.034} $ & $ 49.2 \pm 1.1 $ & \nodata & \nodata & \nodata & CPL \\
XRR100621A & $ -1.72_{-0.13}^{+0.13} $ & \nodata & $ 65.1_{-7.5}^{+13} $ & 32.02/58 & $ 0.931 \pm 0.024 $ & $ 206 \pm 3 $ & \nodata & \nodata & \nodata & CPL \\
XRR101024A & $ -1.09_{-0.32}^{+0.35} $ & \nodata & $ 53.1_{-5.3}^{+7.4} $ & 53.08/58 & $ 0.924 \pm 0.069 $ & $ 14.5 \pm 0.7 $ & \nodata & \nodata & \nodata & CPL \\
XRR110411A & $ -1.51_{-0.32}^{+0.35} $ & \nodata & $ 37.3_{-11}^{+6.3} $ & 45.33/58 & $ 1.09 \pm 0.08 $ & $ 33.0 \pm 1.6 $ & \nodata & \nodata & \nodata & CPL \\
XRR110726A & $ -0.622_{-0.832}^{+1.023} $ & \nodata & $ 42.8_{-7.7}^{+11.6} $ & 56.53/58 & $ 1.11_{-0.27}^{+0.28} $ & $ 2.18_{-0.30}^{+0.33} $ & \nodata & \nodata & \nodata & CPL \\
XRR111022A & $ -0.872_{-0.38}^{+0.42} $ & \nodata & $ 56.4_{-6.0}^{+9.9} $ & 58.22/58 & $ 0.877_{-0.075}^{+0.076} $ & $ 19.2 \pm 1.2 $ & \nodata & \nodata & \nodata & CPL \\
XRR120102A & $ -1.49_{-0.04}^{+0.03} $ & \nodata & $ 967_{-399}^{+825} $ & 162.98/118 & $ 0.752 \pm 0.035 $ & $ 43.2 \pm 1.2 $ & $ 1.29_{-0.06}^{+0.07}$  & $ 1.00_{-0.06}^{+0.07} $  & $ 2.40_{-0.87}^{+1.65} $ & joint-CPL \\
XRR120326A & $ -1.41_{-0.32}^{+0.35} $ & \nodata & $ 41.1_{-7.9}^{+6.0} $ & 55.90/58 & $ 1.05 \pm 0.08 $ & $ 25.2 \pm 1.3 $ & \nodata & \nodata & \nodata & CPL \\
XRR120703A & $ -1.34_{-0.16}^{+0.14} $ & \nodata & $ 168_{-44}^{+116} $ & 132.04/118 & $ 0.754 \pm 0.050 $ & $ 37.9 \pm 1.6 $ & $ 1.26_{-0.12}^{+0.13}$  & $ 1.26 \pm 0.11 $  & < 74.7 & joint-CPL \\
XRR120802A & $ -1.10_{-0.46}^{+0.52} $ & \nodata & $ 49.9_{-7.0}^{+11} $ & 58.14/58 & $ 0.953_{-0.107}^{+0.109} $ & $ 16.4 \pm 1.3 $ & \nodata & \nodata & \nodata & CPL \\
XRR120811C & $ -1.40_{-0.29}^{+0.31} $ & \nodata & $ 42.7_{-6.2}^{+5.2} $ & 48.95/58 & $ 1.05 \pm 0.06 $ & $ 28.5 \pm 1.1 $ & \nodata & \nodata & \nodata & CPL \\
XRR120927A & $ -0.556_{-0.38}^{+0.42} $ & \nodata & $ 59.5_{-5.4}^{+8.3} $ & 57.34/58 & $ 0.812_{-0.069}^{+0.069} $ & $ 23.1 \pm 1.4 $ & \nodata & \nodata & \nodata & CPL \\
XRR121123A & $ -0.917_{-0.21}^{+0.22} $ & \nodata & $ 60.8_{-3.7}^{+5.0} $ & 57.43/58 & $ 0.840_{-0.034}^{+0.034} $ & $ 124 \pm 4 $ & \nodata & \nodata & \nodata & CPL \\
XRR121128A & $ -1.02_{-0.33}^{+0.45} $ & $ -2.33_{-0.76}^{+0.17} $ & $ 50.9_{-6.1}^{+6.9} $ & 33.21/57 & $ 0.912_{-0.034}^{+0.034} $ & $ 58.3 \pm 1.5 $ & \nodata & \nodata & \nodata & Band \\
XRR130627A & $ 0.32_{-1.24}^{+1.69} $ & $ -2.23_{-0.61}^{+0.40} $ & $ 53.6_{-26}^{+30} $ & 60.28/57 & $ 0.842_{-0.246}^{+0.244} $ & $ 7.66_{-1.44}^{+1.46} $ & \nodata & \nodata & \nodata & Band \\
XRR130701A & $ -1.11_{-0.36}^{+0.39} $ & \nodata & $ 88.7_{-18}^{+83} $ & 59.58/58 & $ 0.734_{-0.036}^{+0.036} $ & $ 43.8 \pm 1.3 $ & \nodata & \nodata & \nodata & CPL \\
XRR130727A & $ -1.14_{-0.34}^{+0.56} $ & $ -1.89_{-0.25}^{+0.12} $ & $ 71.8_{-23}^{+47} $ & 54.12/57 & $ 0.786_{-0.033}^{+0.033} $ & $ 38.4 \pm 1.0 $ & \nodata & \nodata & \nodata & Band \\
XRR130925A & $ -1.75_{-0.17}^{+0.17} $ & \nodata & $ 36.0_{-14}^{+7.2} $ & 45.25/58 & $ 1.04 \pm 0.03 $ & $ 403 \pm 7 $ & \nodata & \nodata & \nodata & CPL \\
XRR140108A & $ -1.32_{-0.14}^{+0.12} $ & \nodata & $ 122_{-20}^{+36} $ & 135.82/112 & $ 0.720 \pm 0.034 $ & $ 71.4 \pm 1.8 $ & $ 1.43_{-0.11}^{+0.11}$  & $ 1.44_{-0.11}^{+0.12} $  & \nodata & joint-CPL \\ \hline
GRB080714 & $ -1.11_{-0.25}^{+0.43} $ & $-1.95_{-0.31}^{+0.18}$ & $ 109_{-39.0}^{+70.5} $ & 84.08/85 & $ 0.678 \pm 0.047 $ & $ 25.5_{-1.1}^{+1.2} $ & $ 1.55_{-0.15}^{+0.16}$  & $ 1.26 \pm 0.13 $  & $3.02_{-1.77}^{+3.71}$ & joint-Band \\
GRB080804 & $ -0.56_{-0.18}^{+0.16} $ & \nodata & $ 192_{-29}^{+42} $ & 93.80/81 & $ 0.531 \pm 0.050 $ & $ 37.9 \pm 2.1 $ & $ 1.17_{-0.09}^{+0.10}$  & $ 1.10_{-0.09}^{+0.10} $  & \nodata & joint-CPL \\
GRB080916A & $ -1.00_{-0.19}^{+0.38} $ & $-2.06_{-1.65}^{+0.22}$ & $ 92.4_{-28.7}^{+29.0} $ & 119.93/117 & $ 0.695 \pm 0.034 $ & $ 42.2 \pm 1.3 $ & $ 1.24 \pm 0.09 $  & $1.29 \pm 0.08 $ & $< 6.02$ & joint-Band \\
GRB081121 & $ -0.54_{-0.16}^{+0.18} $ & $ -2.14_{-0.27}^{+0.17} $ & $ 175_{-26}^{+32} $ & 132.76/125 & $ 0.531 \pm 0.059 $ & $ 50.8 \pm 3.3 $ & $ 1.24_{-0.09}^{+0.10}$  & $ 1.10_{-0.09}^{+0.10} $ & $ 1.59_{-0.40}^{+0.54} $ & joint-Band \\
GRB081222 & $ -0.99_{-0.14}^{+0.21} $ & $-1.97_{-0.32}^{+0.18}$ & $ 130_{-32.5}^{+35.7} $ & 105.21/81 & $ 0.664 \pm 0.022 $ & $ 52.2 \pm 1.1 $ & $ 1.37 \pm 0.06$  & $ 1.15 \pm 0.07 $  & $3.08_{-1.88}^{+6.01}$ & joint-Band \\
GRB090102 & $ -0.99 \pm 0.04 $ & \nodata & $ 488_{-48}^{+58} $ & 246.83/188 & $ 0.627 \pm 0.051 $ & $ 70.6 \pm 3.7 $ & $ 1.24_{-0.06}^{+0.07}$  & $ 1.28 \pm 0.07 $  & $ 1.29_{-0.17}^{+0.19} $ & joint-CPL \\
GRB090424 & $ -1.11_{-0.06}^{+0.07} $ & $ -2.22_{-0.41}^{+0.16} $ & $ 153_{-18}^{+21} $ & 155.41/85 & $ 0.709 \pm 0.024 $ & $ 218 \pm 4 $ & $ 1.18 \pm 0.03$  & $ 1.15 \pm 0.03$ & $ 1.18_{-0.58}^{+0.96}$ & joint-Band \\
GRB090926B & $ -0.43_{-0.24}^{+0.26} $ & \nodata & $ 71.3_{-4.5}^{+6.1} $ & 50.77/58 & $ 0.704 \pm 0.035 $ & $ 71.2 \pm 2.4 $ & \nodata & \nodata & \nodata & CPL \\
GRB100816A & $ -0.45_{-0.14}^{+0.15} $ & $ -2.22_{-0.43}^{+0.22} $ & $ 131_{-14}^{+18} $ & 163.25/125 & $ 0.548 \pm 0.033 $ & $ 19.5 \pm 0.7 $ & $ 1.17 \pm 0.07 $  & $ 1.14 \pm 0.07 $ & $ 1.23_{-0.60}^{+0.93} $ & joint-Band \\
GRB110625A & $ -1.22 \pm 0.04 $ & \nodata & $ 219_{-16}^{+19} $ & 221.12/114 & $ 0.675 \pm 0.027 $ & $ 281 \pm 7 $ & $ 1.19 \pm 0.04 $  & $ 1.33 \pm 0.04 $  & $2.98_{-1.62}^{+2.03}$ & joint-CPL \\
GRB110731A & $ -1.21 \pm 0.06 $ & \nodata & $ 548_{-123}^{+225} $ & 162.41/118 & $ 0.554 \pm 0.022 $ & $ 59.4 \pm 1.4 $ & $ 2.10 \pm 0.10$  & $ 2.20 \pm 0.12 $  & $1.41_{-0.98}^{+1.36}$ & joint-CPL \\
GRB121011A & $ -1.01_{-0.28}^{+0.37} $ & $-2.37_{-2.06}^{+0.42}$ & $ 112_{-30}^{+57} $ & 118.51/85 & $ 0.555 \pm 0.084 $ & $ 22.6 \pm 2.1 $ & $ 1.46_{-0.19}^{+0.22}$  & $ 1.50_{-0.22}^{+0.24} $  & $23.1_{-21.9}^{+310.9}$ & joint-Band \\
GRB131229A & $ -0.85_{-0.04}^{+0.05} $ & $-2.46_{-0.63}^{+0.27}$ & $ 401_{-47}^{+56} $ & 201.06/187 &  $ 0.526 \pm 0.020 $ & $ 68.9 \pm 1.5 $ & $ 1.23 \pm 0.04$  & $ 1.25 \pm 0.04 $  & $ 1.33_{-0.18}^{+0.21} $ & joint-Band \\
\enddata
 \tablenotetext{a}{Low-energy spectral index}
 \tablenotetext{b}{High-energy spectral index}
 \tablenotetext{c}{Fluence ratio of $S_{25-50{\rm keV}}$/$S_{50-100{\rm keV}}$ derived from the best-fit model.}
 \tablenotetext{d}{BAT 15--150-keV energy fluence in $10^{-7}$ erg cm$^{-2}$ s$^{-1}$ derived from the best-fit model.}
 \tablenotetext{e}{Constant factor  relative to the BAT data.}
 \tablenotetext{f}{The best-fit model.}
\end{deluxetable}
\end{longrotatetable}

\begin{deluxetable}{l c c c c}[t]

\tablecaption{\textbf{Summary} of redshift, peak energy ($E_{{\rm peak}}^{{\rm src}}\equiv (1+z)E_{{\rm peak}}^{{\rm obs}}$), and equivalent isotropic energy ($E_{{\rm iso}}$)  of our samples.\label{tbl:intrinsic_prompt_result}}
\tablehead{
  \colhead{Events} & \colhead{Redshift ($z$)} & \colhead{$E_{{\rm peak}}^{{\rm src}}$ [keV]} & \colhead{$E_{{\rm iso}}$ [$10^{52}$ erg]}
}
\startdata
XRF050406 & $2.7_{-0.41}^{+0.29}$ & $106_{-45}^{+23}$ & $<0.342$ \\
XRF050416A & 0.6535 & $ 31.6_{-19}^{+6.3} $ & $ 0.0548_{-0.0067}^{+0.0092} $ \\
XRF050819 & 2.5043 & $ < 74.6 $ &  $ < 0.830 $ \\
XRF050824 & 0.83 & $ < 33.7 $ &  $ < 0.102 $ \\
XRF060926 & 3.208 & $ < 93.0 $ &  $ < 0.741 $ \\
XRF080520 & 1.545 & $ < 49.9 $ &  $ < 0.0714 $ \\
XRF081007 & 0.5295 & $ < 29.8 $ &  $ < 0.212 $ \\
XRF100425A & 1.755 & $ < 53.1 $ &  $ < 0.540 $ \\
XRF110808A & 1.348 & $< 54.0$ & $<0.303$ \\
XRF120724A & 1.48 & $50.6_{-40.4}^{+9.7}$ & $0.662_{-0.214}^{+0.331}$ \\
XRF130612A & 2.006 & $128_{-52}^{+120}$ & $0.934_{-0.215}^{+0.253}$ \\ \hline
XRR050318 & 1.44 & $ 118_{-14}^{+21} $ & $ 1.23_{-0.06}^{+0.06} $ \\
XRR050525A & 0.606 & $ 123_{-4.1}^{+4.8} $ & $ 3.17_{-0.09}^{+0.10} $ \\
XRR060206 & 4.045 & $ 352_{-48}^{+110} $ & $ 1.78_{-0.18}^{+0.25} $ \\
XRR060707 & 3.43 & $ 271_{-38}^{+81} $ & $ 7.57_{-0.66}^{+0.71} $ \\
XRR060927 & 5.6 & $ 286_{-46}^{+120} $ & $ 4.23_{-0.24}^{+0.34} $ \\
XRR061222B & 3.355 & $ 196_{-44}^{+60} $ & $ 10.8_{-0.8}^{+0.8} $ \\
XRR071010B & 0.947 & $ 87.6_{-11.0}^{+11.0} $ & $ 2.19_{-0.21}^{+0.31} $ \\
XRR080207 & 2.0858 & $ 282_{-49}^{+140} $ & $ 13.9_{-1.3}^{+2.2} $ \\
XRR080603B & 2.69 & $ 238_{-27}^{+51} $ & $ 8.21_{-0.79}^{+1.00} $ \\
XRR081221 & 2.26 & $ 214_{-59}^{+93} $ & $ 25.9_{-2.3}^{+3.6} $ \\
XRR090423 & 8.0 & $ 439_{-46}^{+66} $ & $ 8.37_{-0.85}^{+1.50} $ \\
XRR100615A & 1.398 & $ 150_{-17}^{+35} $ & $ 5.39_{-0.54}^{+0.84} $ \\
XRR100621A & 0.542 & $ 100_{-11}^{+20} $ & $ 3.97_{-0.32}^{+0.44} $ \\
XRR110726A & 1.036 & $ 87.2_{-15.6}^{+23.6} $ & $ 0.148_{-0.077}^{+0.179} $ \\
XRR120326A & 1.798 & $ 114_{-22}^{+16} $ & $ 4.66_{-0.17}^{+0.17} $ \\
XRR120802A & 3.796 & $ 239_{-33}^{+54} $ & $ 3.12_{-0.18}^{+0.18} $ \\
XRR120811C & 2.671 & $ 156_{-22}^{+18} $ & $ 8.88_{-1.10}^{+1.90} $ \\
XRR121128A & 2.20 & $ 162_{-19}^{+21} $ & $ 13.0_{-0.94}^{+1.3} $ \\
XRR130701A & 1.155 & $ 191_{-40}^{+180} $ & $ 3.50_{-0.45}^{+0.94} $ \\ 
XRR130925A & 0.35 & $ 48.6_{-19}^{+9.8} $ & $ 3.11_{-0.42}^{+0.50} $ \\ \hline
GRB080804 & 2.2045 & $ 523_{-86}^{+120} $ & $ 12.2_{-1.3}^{+1.6} $ \\
GRB080916A & 0.689 & $ 173_{-25}^{+38} $ & $ 1.21_{-0.08}^{+0.10} $ \\ 
GRB081121 & 2.512 & $ 614_{-92}^{+114} $ & $ 24.4_{-3.4}^{+3.7} $ \\
GRB081222 & 2.77 & $ 490_{-124}^{+135} $ & $ 27.0_{-5.0}^{+5.1} $ \\
GRB090102 & 1.547 & $ 1180_{-80}^{+250} $ & $ 22.2_{-0.6}^{+0.9} $ \\
GRB090424 & 0.544 & $ 237_{-29}^{+34} $ & $ 4.79_{-0.79}^{+0.71} $ \\
GRB090926B & 1.24 & $ 155_{-15}^{+17} $ & $ 4.46_{-1.25}^{+1.82} $ \\
GRB100816A & 0.8034 & $ 239_{-25}^{+28} $ & $ 0.714_{-0.158}^{+0.209} $ \\
GRB110731A & 2.83 & $ 1910_{-570}^{+830} $ & $ 44.0_{-2.7}^{+3.0} $ \\ \hline
\enddata
\end{deluxetable}

\section{Results of the afterglow emissions} \label{sec:afterglow}
\subsection{Energy flux light-curves of X-ray afterglows} \label{ss:xrt_Xray_afterglow_result}
\begin{figure}[h]
\centering
\includegraphics[width=6cm,angle=270]{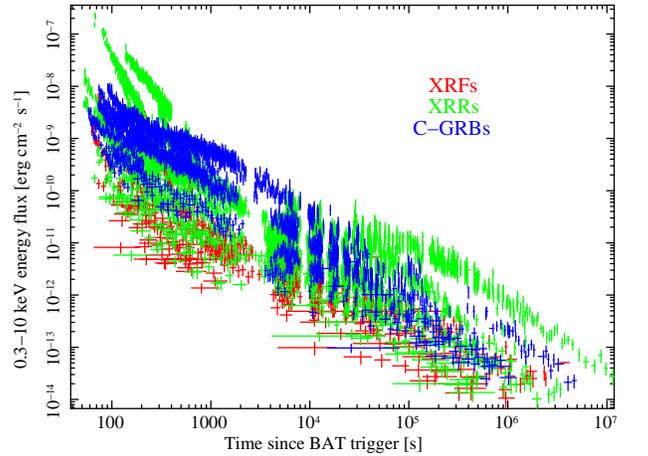}
\figcaption{Energy-flux light curves of the X-Ray afterglows  of all our samples.\label{fig:sum_Xray_afterglow_flux}}
\end{figure}

\begin{figure*}[h]
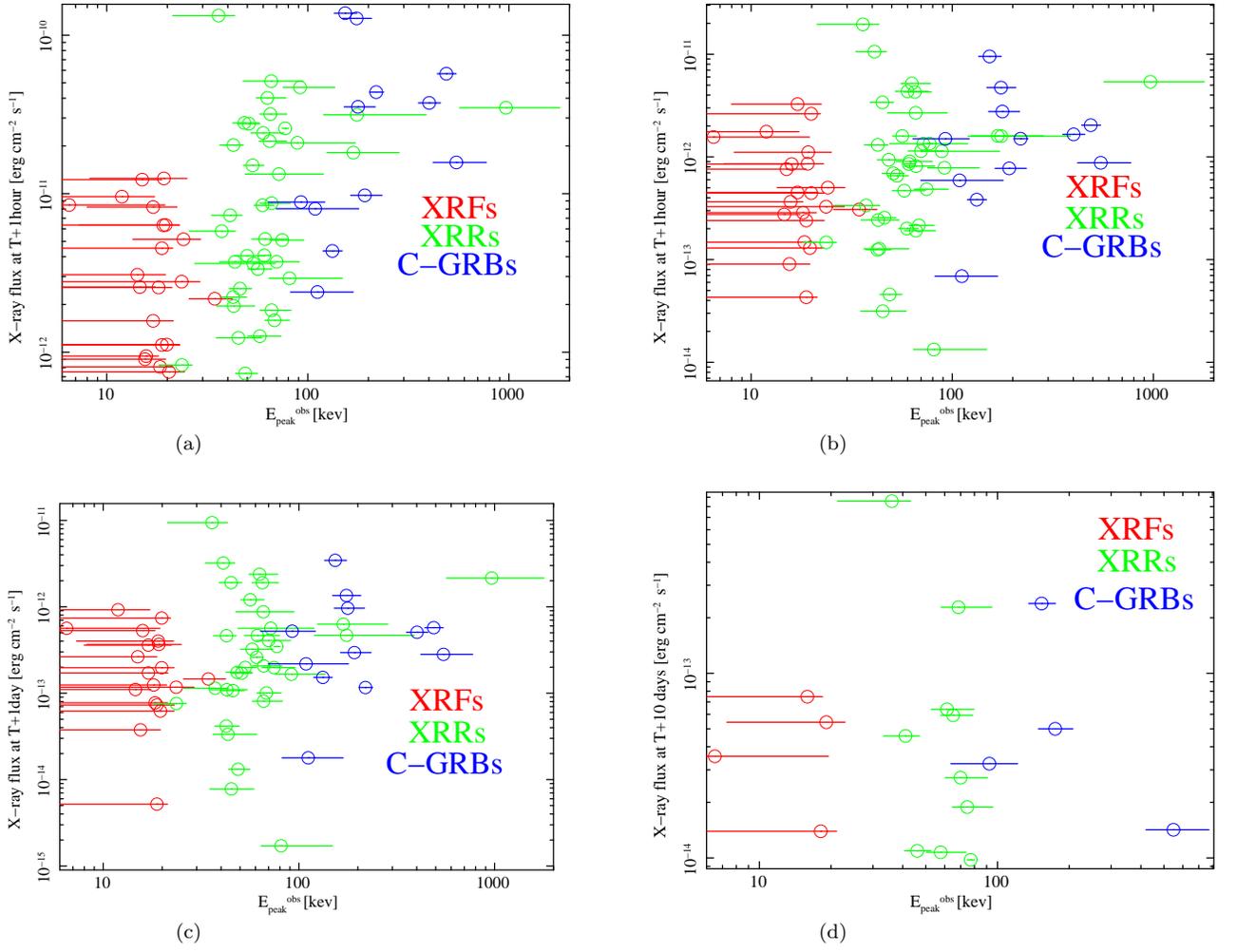

\gridline{\rotatefig{270}{ep_afterglow_1hr_flux_v2.eps}{0.33\textwidth}{(a)}
  \rotatefig{270}{ep_afterglow_10hr_flux_v2.eps}{0.33\textwidth}{(b)}
}
\gridline{\rotatefig{270}{ep_afterglow_1day_flux_v2.eps}{0.33\textwidth}{(c)}
          \rotatefig{270}{ep_afterglow_10days_flux_v2.eps}{0.33\textwidth}{(d)}
          }
\caption{Energy flux at (a) 1 hour, (b) 10 hours, (c) 1 day, and (d) 10 days  after the  trigger time of \textsl{Swift}/BAT versus $E_{peak}^{obs}$.\label{fig:Xray_afterglow_flux_timing}}
\end{figure*}

\begin{figure*}[h]
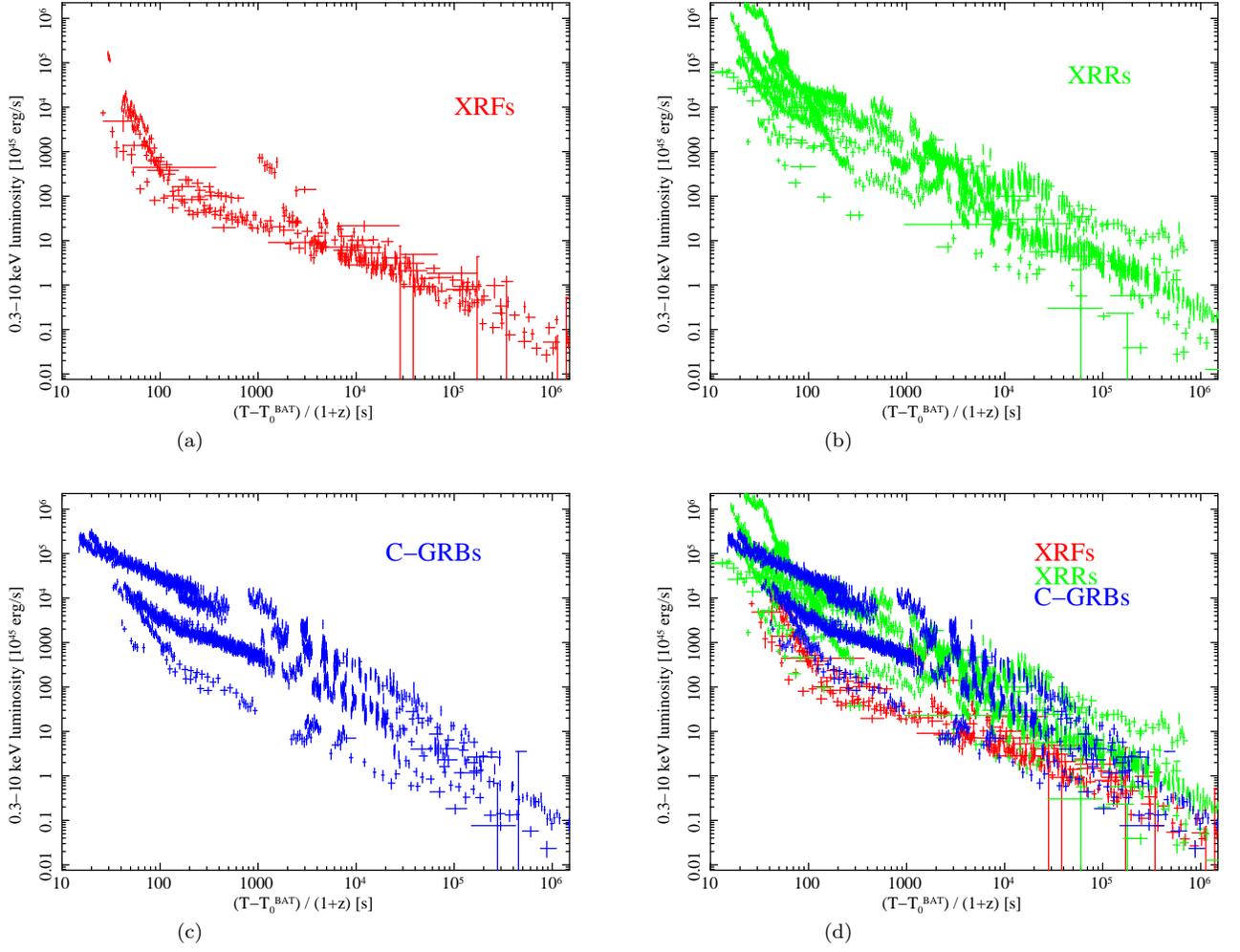

\gridline{\rotatefig{270}{xrf_xrt_lum.eps}{0.33\textwidth}{(a)}
          \rotatefig{270}{xrr_xrt_lum.eps}{0.33\textwidth}{(b)}
           }
\gridline{\rotatefig{270}{cgrb_xrt_lum.eps}{0.33\textwidth}{(c)}
          \rotatefig{270}{sum_xrt_lum.eps}{0.33\textwidth}{(d)}
          }
\caption{ X-ray luminosity light-curves of (a) XRFs, (a) XRRs, (a) C-GRBs, (a) and  all our samples.\label{fig:sum_Xray_afterglow_lumi}}
\end{figure*}

  The energy-flux light curves of the X-ray afterglows in the 0.3--10 keV band  of our sample observed by \textsl{Swift}/XRT  are plotted in figure \ref{fig:sum_Xray_afterglow_flux}. The  energy flux of the XRFs  has a tendency to be  slightly  lower than  those of the XRRs and C-GRBs. Table~\ref{tbl:Xray_afterglow_flux_fitting_results} summarizes the results of the light-curve model-fitting. Figure \ref{fig:Xray_afterglow_flux_timing} shows $E_{peak}^{obs}$ versus energy flux at 1 hour, 10 hours, 1 day, and 10 days after the trigger time of \textsl{Swift}/BAT. 
  These results indicate that the afterglows of the XRFs tend to be fainter than that of the C-GRBs between $10^3$--$10^4$ s after the trigger time and that the tendency disappears as time elapses.  We calculated the X-ray luminosity in the 0.3--10 keV band, using equation (\ref{eq:xray_lumi}), for the events with known redshifts  in our samples (11 XRFs, 20 XRRs, and 9 C-GRBs),  and  summarized the result and  X-ray  luminosity light-curves in tables \ref{tbl:XrayLumFitResult} and \ref{tbl:XrayLumResult2} and figure \ref{fig:sum_Xray_afterglow_lumi}, respectively.    The above-mentioned  trend in the X-ray afterglows of the XRFs  is more  pronounced in this figure of   the energy fluxes.  Furthermore, we found that  if the steep decay phase ($\Gamma_1 < -2$) is ignored,  the X-ray luminosities of the XRFs and XRRs decay more  slowly than those of the C-GRBs.

\onecolumngrid

\onecolumngrid

\twocolumngrid

\begin{longrotatetable}
\begin{deluxetable}{l c c c c c c c c c}
\tablecaption{Fitting results of the X-ray afterglow light-curves (0.3--10 keV energy flux).\label{tbl:Xray_afterglow_flux_fitting_results}}
\tablehead{
  \colhead{Events} & \colhead{$\Gamma_1$\tablenotemark{a}} & \colhead{$t_1$\tablenotemark{b} [sec]} & \colhead{$\Gamma_2$\tablenotemark{c}} & \colhead{$t_2$\tablenotemark{d} [sec]} & \colhead{$\Gamma_3$\tablenotemark{e}} & \colhead{$t_3$\tablenotemark{f} [sec]} & \colhead{$\Gamma_4$\tablenotemark{g}} & \colhead{$\chi^2$/d.o.f.} & \colhead{Best-fit model\tablenotemark{h}}
  }
\startdata
XRF050406 & $ -2.78_{-0.20}^{+0.23} $ & $ 917 \pm 291 $ & $ -0.504_{-0.128}^{+0.126} $ & \nodata & \nodata & \nodata & \nodata & 1.84/5 & brkpow \\
XRF050416A & $ -1.90_{-0.27}^{+0.32} $ & $ 184 \pm 5 $ & $ -0.363 \pm 0.021 $ & $ 1340_{-110}^{+130} $ & $ -0.868 \pm 0.013 $ & \nodata & \nodata & 85.7/94 & brkpow2 \\
XRF050819 & $ -3.87 \pm +0.19 $ & $ 441_{-23}^{+26} $ & $ -0.738_{-0.120}^{+0.080} $ & \nodata & \nodata & \nodata & \nodata & 30.4/18 & brkpow \\
XRF050824 & $ -0.390_{-0.087}^{+0.069} $ & $ 6.47_{-0.91}^{+1.50} \times 10^4$ & $ -0.850_{-0.052}^{+0.048} $ & \nodata & \nodata & \nodata & \nodata & 49.0/37 & brkpow \\
XRF060219 & $ -5.92_{-0.57}^{+0.51} $ & $ 222_{-10}^{+8} $ & $ -0.553 \pm 0.043 $ & $ 2.62_{-0.59}^{+0.57} \times 10^4$ & $ -1.36_{-0.13}^{+0.10} $ & \nodata & \nodata & 6.98/20 & brkpow2 \\
XRF060428B & $ -4.54_{-0.16}^{+0.12} $ & $ 666_{-40}^{+34} $ & $ -0.951_{-0.048}^{+0.039} $ & \nodata & \nodata & \nodata & \nodata & 133/129 & brkpow \\
XRF060923B & $ -0.612_{-0.080}^{+0.085} $ & $ 5910_{-970}^{+960} $ & $ -2.41_{-0.47}^{+0.34} $ & \nodata & \nodata & \nodata & \nodata & 10.1/11 & brkpow \\
XRF060926 & $ -1.45_{-0.26}^{+0.21} $ & \nodata & \nodata & \nodata & \nodata & \nodata & \nodata & 3.86/10 & PL \\
XRF070330 & $ -0.849 \pm 0.075 $ & \nodata &  \nodata & \nodata & \nodata & \nodata & \nodata & 17.0/17 & PL \\
XRF070714A & $ -0.537_{-0.019}^{+0.017} $ & $ 310_{-28}^{+24} $ & $ -0.673_{-0.087}^{+0.073} $ & $ 5650_{-3500}^{+16000} $ & $ -1.04_{-0.31}^{+0.15} $ & \nodata & \nodata & 8.47/8 & brkpow2 \\
XRF080218B & $ -0.929_{-0.058}^{+0.057} $ & $ 6.59_{-1.90}^{+2.70} \times 10^4$ & $ -1.72_{-0.43}^{+0.28} $ & \nodata & \nodata & \nodata & \nodata & 6.66/17 & brkpow \\
XRF080520 & $ -1.00_{-0.09}^{+0.08} $ & \nodata & \nodata & \nodata & \nodata & \nodata & \nodata & 5.05/9 & PL \\
XRF081007 & $ -4.02_{-0.13}^{+0.13} $ & $ 220 \pm 5 $ & $ -0.735_{-0.017}^{+0.016} $ & $ 3.84_{-0.79}^{+0.88} \times 10^4$ & $ -1.20_{-0.06}^{+0.05} $ & \nodata & \nodata & 63.9/78 & brkpow2 \\
XRF100425A & $ -4.54_{-0.08}^{+0.09} $ & $ 337_{-10}^{+7.3} $ & $ -0.543_{-0.030}^{+0.029} $ & $ 4.35_{-1.10}^{+1.20} \times 10^4$ & $ -1.26_{-0.19}^{+0.14} $ & \nodata & \nodata & 26.9/24 & brkpow2 \\
XRF110319A & $ -4.13_{-0.13}^{+0.14} $ & $ 142 \pm 4 $ & $ -0.611_{-0.021}^{+0.021} $ & $ 7670_{-850}^{+890} $ & $ -1.26_{-0.07}^{+0.06} $ & \nodata & \nodata & 64.2/64 & brkpow2 \\
XRF110808A & $ -3.57_{-0.09}^{+0.08} $ & $ 564_{-20}^{+23} $ & $ -0.400_{-0.034}^{+0.030} $ & $ 4.35_{-1.90}^{+1.20} \times 10^4$ & $ -1.07_{-0.11}^{+0.09} $ & \nodata & \nodata & 28.4/30 & brkpow2 \\
XRF111129A & $ -0.507_{-0.055}^{+0.058} $ & $ 3040_{-570}^{+660} $ & $ -1.21 \pm 0.06 $ & \nodata & \nodata & \nodata & \nodata & 47.6/46 & brkpow \\
XRF120116A & $ -2.83 \pm 0.10 $ & $ 244 \pm 11 $ & $ -0.402_{-0.034}^{+0.033} $ & $ 3.69_{-0.32}^{+0.31} \times 10^4 $ & $ -2.58_{-0.31}^{+0.26} $ & \nodata & \nodata & 25.1/26 & brkpow2 \\
XRF120724A & $ -3.95 \pm 0.16 $ & $ 271_{-15}^{+17} $ & $ -1.66_{-0.09}^{+0.07} $ & $1810_{-160}^{+210}$ & $ -0.154_{-0.082}^{+0.069}$ & \nodata & \nodata & 9.4/17 & brkpow2 \\
XRF120816A & $ -1.07 \pm 0.08 $ & $ 5700_{-720}^{+1420} $ & $ 0.056_{-0.421}^{+0.344} $ & \nodata & \nodata & \nodata & \nodata & 28.5/22 & brkpow \\
XRF121108A & $ -1.50_{-0.19}^{+0.28} $ & $ 658_{-301}^{+284} $ & $ -0.845_{-0.105}^{+0.093} $ & $ 1.64_{-0.43}^{+1.16} \times 10^4$ & $ -1.75_{-0.22}^{+0.16} $ & \nodata & \nodata & 20.2/22 & brkpow2 \\
XRF121212A & $ -5.74_{-0.44}^{+0.51} $ & $ 936_{-31}^{+28} $ & $ -0.736 \pm 0.039 $ & \nodata & \nodata & \nodata & \nodata & 48.9/43 & brkpow \\
XRF130608A & $ -4.68_{-0.27}^{+0.24} $ & $ 616_{-54}^{+65} $ & $ -0.412_{-0.102}^{+0.109} $ & \nodata & \nodata & \nodata & \nodata & 30.5/30 & brkpow \\
XRF130612A & $ -0.466_{-0.076}^{+0.099} $ & $ 1280_{-1020}^{+1040} $ & $ -1.01_{-0.07}^{+0.06} $ & \nodata & \nodata & \nodata & \nodata & 1.9/9 & brkpow \\ 
XRF130812A & $ -0.687_{-0.190}^{+0.280} $ & $ 658_{-301}^{+284} $ & $ -0.845_{-0.105}^{+0.093} $ & \nodata & \nodata & \nodata & \nodata & 33.9/41 & brkpow \\ 
XRF140103A & $ -0.137_{-0.082}^{+0.085} $ & $ 3530_{-280}^{+310} $ & $ -2.08_{-0.21}^{+0.15} $ & \nodata & \nodata & \nodata & \nodata & 42.3/35 & brkpow \\ \hline
XRR050318 & $ -1.09_{-0.07}^{+0.13} $ & $ 1.25_{-0.38}^{+0.25} \times 10^4$ & $ -1.92_{-0.14}^{+0.16} $ & \nodata & \nodata & \nodata & \nodata & 65.4/79 & brkpow \\
XRR050410 & $ -1.02_{-0.06}^{+0.05} $ & \nodata & \nodata & \nodata & \nodata & \nodata & \nodata & 13.1/14 & PL \\
XRR050525A & $ -0.711_{-0.049}^{+0.054} $ & $ 7510_{-1600}^{+1300} $ & $ -1.55 \pm 0.08 $ & \nodata & \nodata & \nodata & \nodata & 26.8/32 & brkpow \\
XRR050915B & $ -2.09_{-0.23}^{+0.26} $ & $ 956_{-140}^{+110} $ & $ -0.429_{-0.033}^{+0.030} $ & $ 8.96_{-2.20}^{+2.10} \times 10^5 $ & $ -1.50_{-0.27}^{+0.20} $ & \nodata & \nodata & 23.3/17 & brkpow2 \\
XRR060206 & $ -0.925_{-0.039}^{+0.095} $ & $ 1700_{-650}^{+820} $ & $ -0.429_{-0.027}^{+0.031} $ & $ 2.78_{-0.38}^{+0.21} \times 10^4$ & $ -1.18_{-0.08}^{+0.09} $ & \nodata & \nodata & 26.4/28 & brkpow2 \\
XRR060707 & $ -1.75_{-0.12}^{+0.13} $ & $ 419 \pm 110 $ & $ -0.758_{-0.033}^{+0.029} $ & $ 7.15_{-2.40}^{+4.80} \times 10^5$ & $ -2.01_{-0.48}^{+0.46} $  & \nodata & \nodata & 36.7/44 & brkpow2 \\
XRR060825 & $ -0.983_{-0.045}^{+0.048} $ & \nodata & \nodata & \nodata & \nodata & \nodata & \nodata & 10.5/11 & PL \\
XRR060927 & $ -0.758_{-0.070}^{+0.17} $ & $ 4370_{-2000}^{+1600} $ & $ -1.53_{-0.41}^{+0.21} $ & \nodata & \nodata & \nodata & \nodata & 8.71/15 & brkpow \\
XRR061222B & $ -3.34_{-0.14}^{+0.15} $ & $ 415_{-49}^{+34} $ & $ -1.59_{-0.13}^{+0.01} $ & \nodata & \nodata & \nodata & \nodata & 36.4/38 & brkpow \\
XRR070612B & $ -2.34_{-0.022}^{+0.018} $ & \nodata & \nodata & \nodata & \nodata & \nodata & \nodata & 10.3/6 & PL \\
XRR070721A & $ -2.80_{-0.51}^{+0.31} $ & $ 325_{-40}^{+49} $ & $ -0.752_{-0.057}^{+0.057} $ & \nodata & \nodata & \nodata & \nodata & 4.77/16 & brkpow \\
XRR071010B & $ -0.663_{-0.063}^{+0.064} $ & \nodata & \nodata & \nodata & \nodata & \nodata & \nodata & 7.09/16 & PL \\
XRR080207 & $ 0.190 \pm 0.230 $ & $ 342_{-45}^{+58} $ & $ -1.78_{-0.07}^{+0.06} $ & \nodata & \nodata & \nodata & \nodata & 139/144 & brkpow \\
XRR080212 & $ -8.01_{-0.07}^{+0.08} $ & $ 594_{-4}^{+3} $ & $ -0.282_{-0.022}^{+0.019} $ & $ 5890_{-370}^{+390} $ & $ -1.13 \pm 0.05 $ & $ 2.34_{-0.36}^{+0.56} \times 10^4$ & $ -1.56_{-0.12}^{+0.10} $ & 198/198 & brkpow3 \\
XRR080603B & $ -3.45_{-0.22}^{+0.16} $ & $ 151_{-9}^{+8} $ & $ -0.835_{-0.041}^{+0.035} $ & \nodata & \nodata & \nodata & \nodata & 100/96 & brkpow \\
XRR081128 & $ -4.89_{-0.15}^{+0.14} $ & $ 473_{-26}^{+32} $ & $ -0.992_{-0.046}^{+0.043} $ & \nodata & \nodata & \nodata & \nodata & 44.0/47 & brkpow \\
XRR081221 & $ -5.81_{-0.07}^{+0.08} $ & $ 205 \pm 1 $ & $ -0.707_{-0.025}^{+0.029} $ & $ 717_{-41}^{+40} $ & $ -1.28 \pm 0.01 $ & $ 3.54_{-1.40}^{+0.74} \times 10^5$ & $ -3.12_{-0.44}^{+1.60} $ & 433/383 & brkpow3 \\
XRR090423 & $ -5.79_{-0.23}^{+0.27} $ & $ 341_{-3}^{+4} $ & $ 0.001_{-0.100}^{+0.390} $ & $ 5100_{-260}^{+260} $ & $ -1.42_{-0.08}^{+0.07} $ & \nodata & \nodata & 35.2/47 & brkpow2 \\
XRR090429B & $ 0.81_{-0.26}^{+0.39} $ & $ 657_{-88}^{+140} $ & $ -1.25_{-0.07}^{+0.06} $ & \nodata & \nodata & \nodata & \nodata & 17.5/20 & brkpow \\
XRR090531A & $ -0.673_{-0.91}^{+0.87} $ & \nodata & \nodata & \nodata & \nodata & \nodata & \nodata & 2.31/4 & PL \\
XRR090813 & $ -0.237_{-0.067}^{+0.080} $ & $ 445_{-37}^{+42} $ & $ -1.15 \pm 0.02 $ & $ 9340_{-3100}^{+3000} $ & $ -1.40 \pm 0.07 $ & \nodata & \nodata & 314/297 & brkpow2 \\
XRR090912 & $ -0.744_{-0.037}^{+0.037} $ & \nodata & \nodata & \nodata & \nodata & \nodata & \nodata & 55.2/57 & PL \\
XRR100615A & $ -4.23_{-0.45}^{+0.31} $ & $ 192 \pm 2 $ & $ -0.084_{-0.036}^{+0.036} $ & $ 2840_{-170}^{+180} $ & $ -0.889 \pm 0.035 $ & \nodata & \nodata & 89.1/90 & brkpow2 \\
XRR100621A & $ -3.80_{-0.024}^{+0.027} $ & $ 419_{-4}^{+5} $ & $ -0.626_{-0.014}^{+0.013} $ & $ 5790_{-520}^{+720} $ & $ -0.932 \pm 0.018 $ & $ 1.13_{-0.16}^{+0.12} \times 10^5 $ & $ -1.58_{-0.14}^{+0.14} $ & 416/354 & brkpow3 \\
XRR101022A & $ -0.372_{-0.652}^{+0.622} $ & \nodata & \nodata & \nodata & \nodata & \nodata & \nodata & 0.55/3 & PL \\
XRR101024A & $ -1.38_{-0.36}^{+0.47} $ & $ 121_{-3.9}^{+4.7} $ & $ 0.032_{-0.06}^{+0.06} $ & $ 1010_{-47}^{+38} $ & $ -1.36_{-0.073}^{+0.066} $ & \nodata & \nodata & 50.0/55 & brkpow2 \\
XRR110411A & $ -5.79_{-0.52}^{+0.38} $ & $ 244_{-4.1}^{+4.2} $ & $ -0.473_{-0.040}^{+0.039} $ & $ 3560_{-650}^{+930} $ & $ -1.24_{-0.081}^{+0.069} $ & \nodata & \nodata & 53.1/48 & brkpow2 \\ 
XRR110726A & $ -0.853_{-0.051}^{+0.052} $ & \nodata & \nodata & \nodata & \nodata & \nodata & \nodata & 16.7/11 & PL \\
XRR120102A & $ -3.82_{-0.32}^{+0.38} $ & $ 163 \pm 8 $ & $ -0.570 \pm 0.026 $ & $ 1.13_{-0.09}^{+0.06} \times 10^4$ & $ -1.05_{-0.03}^{+0.02} $ & $ 1.74_{-0.39}^{+0.34} \times 10^5$ & $ -1.47_{-0.34}^{+0.23} $ & 155.4/154 & brkpow3 \\
XRR120326A & $ -3.09 \pm 0.03 $ & $ 402 \pm 6 $ & $ -0.139 \pm 0.013 $ & $ 1.14_{-0.07}^{+0.08} \times 10^4$ & $ 0.464_{-0.034}^{+0.036} $ & $ 4.32_{-0.07}^{+0.12} \times 10^4$ & $ -1.85_{-0.06}^{+0.07} $ & 209.6/233 & brkpow3 \\
XRR120703A & $ -2.45_{-0.28}^{+0.30} $ & $ 125 \pm 8 $ & $ -0.621_{-0.028}^{+0.029} $ & $ 3880_{-380}^{+400} $ & $ -1.07 \pm 0.02 $ & $ 3.42_{-0.38}^{+0.56} \times 10^5$ & $ -4.29_{-3.08}^{+1.39} $ & 95.0/80 & brkpow3 \\ 
XRR120802A & $ -2.79_{-0.41}^{+0.28} $ & $ 257_{-40}^{+39} $ & $ -0.372_{-0.055}^{+0.059} $ & \nodata   & \nodata   & \nodata   & \nodata   & 29.0/31 & brkpow \\ 
XRR120811C & $ -3.20_{-0.15}^{+0.12} $ & $ 225 \pm 10 $ & $ -0.474_{-0.057}^{+0.054} $ & $ 2820_{-530}^{+840} $ & $ -1.19 \pm 0.07 $ & \nodata   & \nodata   & 140/106 & brkpow2 \\
XRR120927A & $ -3.33_{-0.01}^{+0.02} $ & $ 185_{-15}^{+13} $ & $ -0.885_{-0.039}^{+0.040} $ & $ 9770_{-1150}^{+1570} $ & $ -2.19_{-0.37}^{+0.28} $ & \nodata   & \nodata   & 25.0/28 & brkpow2 \\ 
XRR121123A & $ -5.39_{-0.04}^{+0.08} $ & $ 1360_{-20}^{+10} $ & $ -0.320_{-0.020}^{+0.023} $ & $ 1.63_{-0.10}^{+0.07} \times 10^4$ & $ -1.36_{-0.06}^{+0.08} $ & \nodata   & \nodata  & 210.0/185 & brkpow2 \\ 
XRR121128A & $ -4.18_{-0.16}^{+0.33} $ & $ 149_{-3}^{+2} $ & $ -0.548_{-0.022}^{+0.038} $ & $ 1500 \pm 90 $ & $ -1.60_{-0.03}^{+0.04} $ & \nodata   & \nodata   & 110.0/105 & brkpow2 \\ 
XRR130627A & $ -0.873_{-0.139}^{+0.125} $ & \nodata   & \nodata   & \nodata  & \nodata   & \nodata   & \nodata   & 14.0/17 & PL \\ 
XRR130701A & $ -2.24_{-0.18}^{+0.19} $ & $ 123 \pm 4 $ & $ -0.785_{-0.048}^{+0.043} $ & $ 471_{-48}^{+53} $ & $ -1.27 \pm 0.02 $ & \nodata   & \nodata   & 110.0/123 & brkpow2 \\ 
XRR130727A & $ -0.994_{-0.051}^{+0.046} $ & \nodata   & \nodata   & \nodata   & \nodata   & \nodata   & \nodata   & 43.0/46 & PL \\ 
XRR130925A & $ -2.43 \pm -0.02 $ & $ 903_{-12}^{+11} $ & $ -0.832 \pm 0.004 $ & $ 3.11_{-0.19}^{+0.25} \times 10^5 $ & $ -1.30_{-0.03}^{+0.02} $ & \nodata   & \nodata   & 900/799 & brkpow2 \\ 
XRR140108A & $ -3.06_{-0.11}^{+0.13} $ & $ 405 \pm 7 $ & $ -0.497 \pm 0.015 $ & $ 7070_{-260}^{+330} $ & $ -1.31_{-0.02}^{+0.03} $ & \nodata   & \nodata   & 340/225 & brkpow2 \\ \hline
GRB080714 & $ -1.13 \pm 0.03 $ & \nodata & \nodata & \nodata & \nodata & \nodata & \nodata & 71.5/43 & PL \\
GRB080804 & $ -1.10_{-0.02}^{+0.01} $ & \nodata & \nodata & \nodata & \nodata & \nodata & \nodata & 82.3/101 & PL \\
GRB080916A & $ -3.41_{-0.14}^{+0.14} $ & $ 319_{-28}^{+33} $ & $ -0.745_{-0.071}^{+0.074} $ & $ 3.17_{-1.20}^{+1.50} \times 10^4$ & $ -1.21_{-0.09}^{+0.08} $ & \nodata & \nodata & 97.0/117 & brkpow2 \\
GRB081121 & $ -1.43 \pm 0.02 $ & \nodata & \nodata & \nodata & \nodata & \nodata & \nodata & 193/147 & PL \\
GRB081222 & $ -0.888_{-0.024}^{+0.024} $ & $ 635_{-110}^{+120} $ & $ -1.11_{-0.02}^{+0.02} $ & $ 7.82_{-2.60}^{+1.70} \times 10^4$ & $ -1.97_{-0.24}^{+0.24} $ & \nodata & \nodata & 492/418 & brkpow2 \\
GRB090102 & $ -0.977_{-0.160}^{+0.171} $ & $ 1460_{-590}^{+990} $ & $ -1.45 \pm 0.04 $ & \nodata & \nodata & \nodata & \nodata & 139/139 & brkpow \\
GRB090424 & $ -0.874_{-0.070}^{+0.065} $ & $ 1540_{-200}^{+140} $ & $ -1.16_{-0.01}^{+0.03} $ & $ 5.92_{-3.90}^{+5.75} \times 10^5$ & $ -1.42_{-0.22}^{+0.20} $ & \nodata & \nodata & 691/663 & brkpow2 \\
GRB090926B & $ -2.25_{-0.12}^{+0.10} $ & $ 660_{-100}^{+130} $ & $ -1.01 \pm 0.12 $ & \nodata & \nodata & \nodata & \nodata & 82.8/99 & brkpow \\
GRB100816A & $ -2.64_{-0.79}^{+0.80} $ & $ 150_{-26}^{+45} $ & $ -1.05 \pm 0.04 $ & \nodata & \nodata & \nodata & \nodata & 43.3/38 & brkpow \\
GRB110610A & $ -2.91_{-0.22}^{+0.28} $ & $ 194_{-29}^{+27} $ & $ -0.130_{-0.348}^{+0.749} $ & $ 930_{-197}^{+579} $ & $ -1.16_{-0.08}^{+0.06} $ & \nodata   & \nodata   & 43.0/50 & brkpow2 \\ 
GRB110625A & $ -2.67_{-0.70}^{+0.56} $ & $ 166_{-10}^{+8} $ & $ -1.12 \pm 0.04 $ & $ 2.32_{-0.41}^{+0.43} \times 10^4$ & $ -2.92_{-0.77}^{+0.45} $ & \nodata   & \nodata   & 53.0/49 & brkpow2 \\ 
GRB110731A & $ -2.83_{-0.67}^{+0.35} $ & $ 94_{-3}^{+4} $ & $ -1.15 \pm 0.02 $ & $ 7100_{-2160}^{+4690} $ & $ -1.30_{-0.05}^{+0.06} $ & \nodata   & \nodata   & 310/301 & brkpow2 \\ 
GRB121011A & $ -1.54_{-0.04}^{+0.03} $ & \nodata   & \nodata   & \nodata   & \nodata   & \nodata   & \nodata   & 24.0/19 & PL \\ 
GRB131229A & $ -1.01 \pm 0.05 $ & $ 424_{-144}^{+277} $ & $ -1.35_{-0.06}^{+0.04} $ & \nodata   & \nodata   & \nodata   & \nodata & 210/226 & brkpow \\ 
\\
\enddata
\tablenotetext{a}{Decay index of the 1st power-law component.}
\tablenotetext{b}{Break time of the 1st component in seconds after the BAT trigger.}
\tablenotetext{c}{Decay index of the 2nd power-law component.}
\tablenotetext{d}{Break time of the 2nd component in seconds after the BAT trigger.}
\tablenotetext{e}{Decay index of the 3rd power-law component.}
\tablenotetext{f}{Break time of the 3rd component in seconds after the BAT trigger.}
\tablenotetext{g}{Decay index of the 4th power-law component.}
\tablenotetext{h}{The models ``brkpow'',  ``brkpow2'', and ``brkpow3''  have two, three, and four decay indices of power-law components, respectively.}
\end{deluxetable}
\end{longrotatetable}

\subsection{X-ray luminosity and the temporal index at 200 seconds after the BAT trigger} \label{ss:x_lum_200}
 Figure \ref{fig:Xray_afterglow_lumi_timing200} shows the plots of  X-ray luminosity versus $E_{{\rm peak}}^{{\rm src}}$, temporal decay index versus $E_{{\rm peak}}^{{\rm src}}$, X-ray luminosity versus $E_{{\rm iso}}$, and temporal decay index versus $E_{{\rm iso}}$. The X-ray luminosity and the temporal decay index are derived at 200 seconds after the \textsl{Swift}/BAT trigger time in the rest frame of the GRBs. The number of the samples is 38.  The correlation coefficients $\sigma$ for the four plots are $\sigma=0.33\pm0.08$, $-0.38_{-0.06}^{+0.03}$, $0.54_{-0.06}^{+0.07}$, and $-0.47\pm0.02$, respectively.  Therefore, $E_{{\rm peak}}^{{\rm src}}$ and $E_{{\rm iso}}$, both of which were derived from the spectrum of the prompt emission, are moderately correlated with the X-ray luminosity and the temporal decay index in the afterglow emission.  In consequence, we confirmed that X-ray afterglow luminosity of the GRBs with a lower $E_{{\rm peak}}^{{\rm src}}$, i.e., softer GRBs, tend to be dimmer and to decay more  slowly than harder GRBs.  Note that we had excluded the data of XRR050318 and XRR071010B because we could not estimate  their luminosity at  the epoch due to  lack of the BAT data.

\subsection{ X-ray luminosity and the temporal index at 10 hours after the BAT trigger} \label{ss:x_lum_36000}
\begin{figure*}[h]
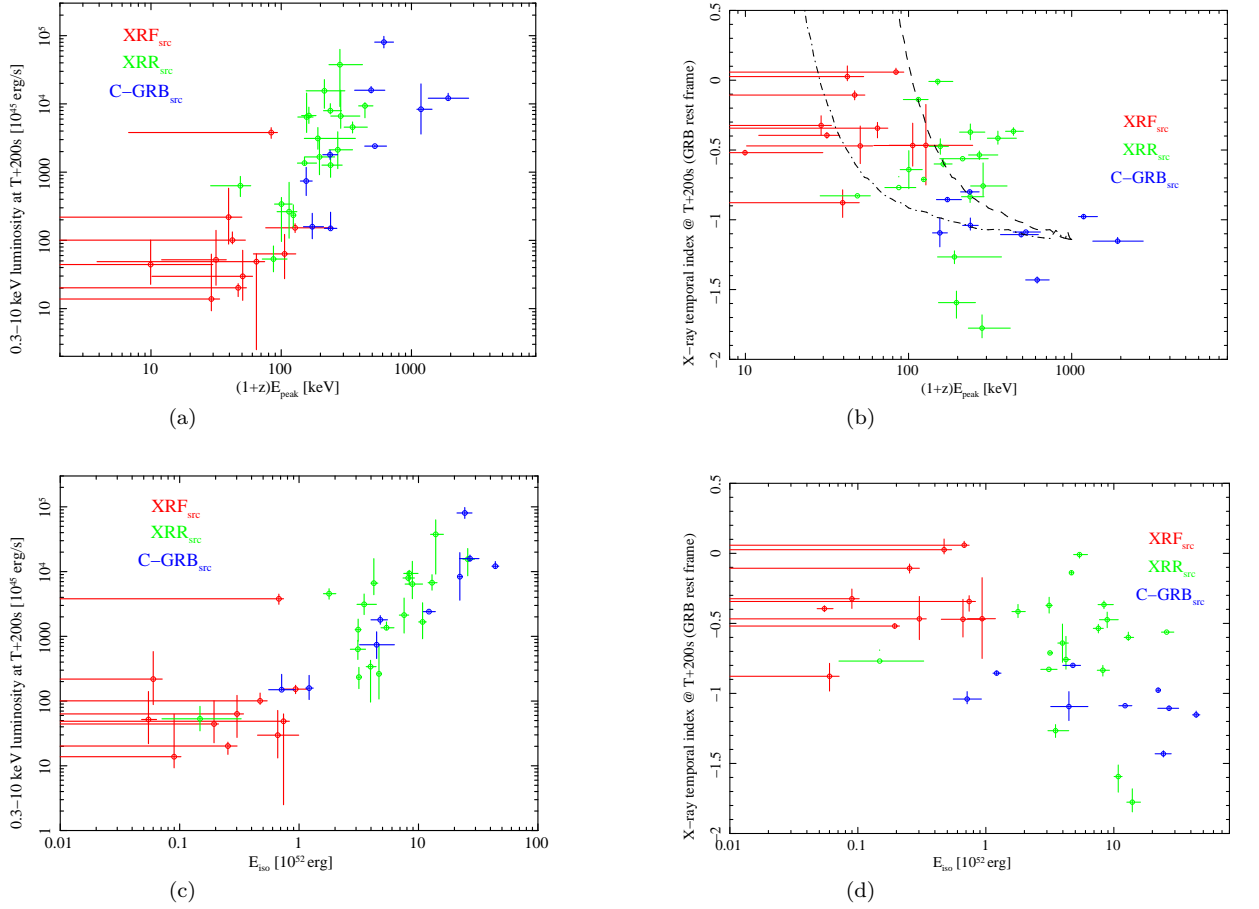

\gridline{\rotatefig{270}{ep_src_afterglow_luminosity_200_v2.eps}{0.30\textwidth}{(a)}
          \rotatefig{270}{ep_src_afterglow_gamma_200_v3.eps}{0.30\textwidth}{(b)}
           }
\gridline{\rotatefig{270}{eiso_afterglow_luminosity_200_v2.eps}{0.30\textwidth}{(c)}
          \rotatefig{270}{eiso_afterglow_gamma_200_v2.eps}{0.30\textwidth}{(d)}
          }
\caption{ Relations between the prompt emission and  early afterglow. (a)X-ray luminosity versus $E_{{\rm peak}}^{{\rm src}}$. (b) Temporal decay index versus $E_{{\rm peak}}^{{\rm src}}$. The dashed and dot-dashed lines  are the  indices derived from the boxfit light-curves with the assumed  $\Gamma$ of 100 and 1000, respectively,  and $(1+z)E_{{\rm peak}}^{{\rm on}}=1000$ keV,  where $\theta_{{\rm obs}}$ is allowed to vary for an range of  0--0.01 rad ($\approx0.6^\circ$). (c) X-ray luminosity versus $E_{{\rm iso}}$. (d) Temporal decay index versus $E_{{\rm iso}}$. These afterglow parameters  are derived  200 seconds after the \textsl{Swift}/BAT trigger time in the rest frame of the GRBs.\label{fig:Xray_afterglow_lumi_timing200}}
\end{figure*}

\begin{figure*}[h]
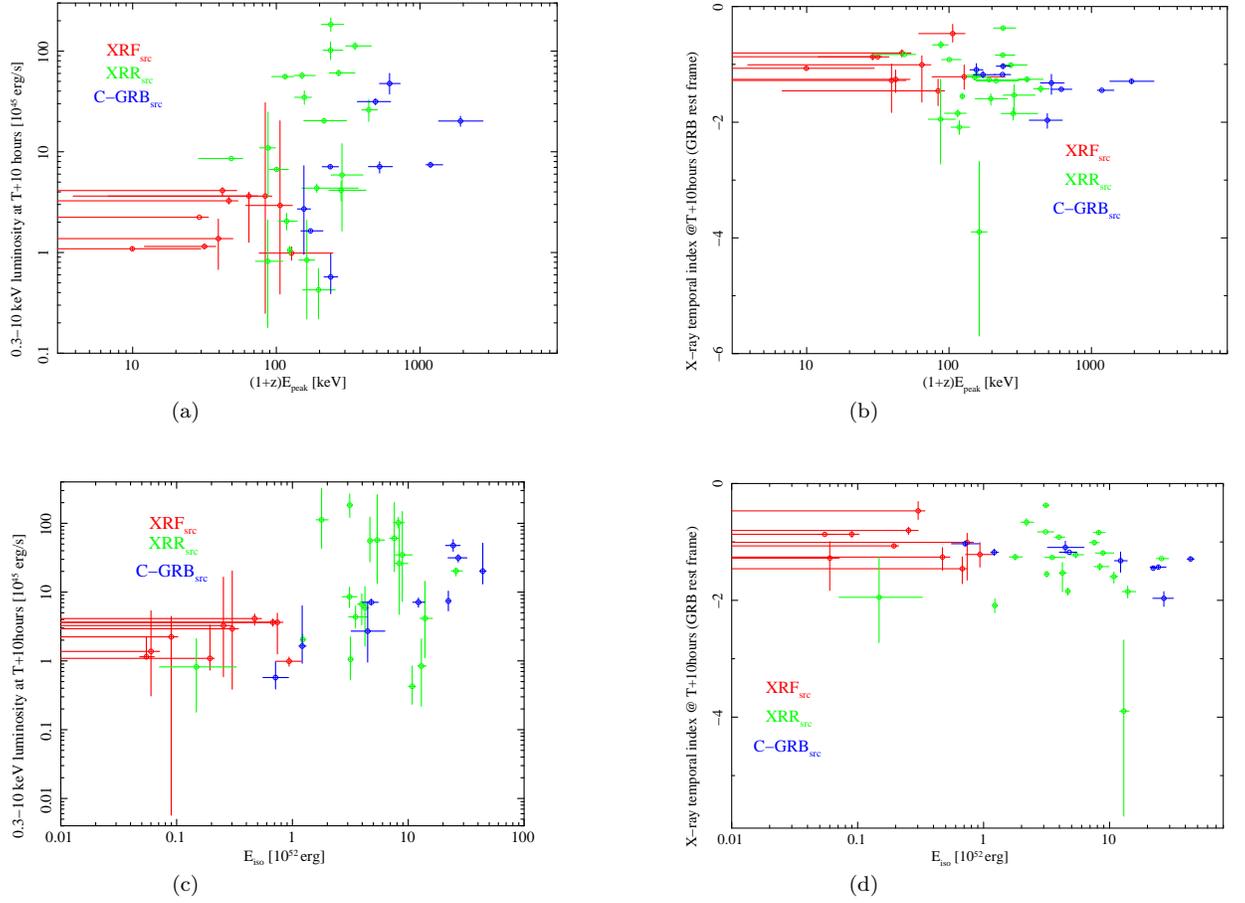

\gridline{\rotatefig{270}{ep_src_afterglow_luminosity_36000_v2.eps}{0.30\textwidth}{(a)}
          \rotatefig{270}{ep_src_afterglow_gamma_36000_v2.eps}{0.30\textwidth}{(b)}
           }
\gridline{\rotatefig{270}{eiso_afterglow_luminosity_36000_v2.eps}{0.30\textwidth}{(c)}
          \rotatefig{270}{eiso_afterglow_gamma_36000_v2.eps}{0.30\textwidth}{(d)}
          }
\caption{    Same as figure~\ref{fig:Xray_afterglow_lumi_timing200} but for the parameters of the afterglow 36000 seconds after the \textsl{Swift}/BAT trigger time.\label{fig:Xray_afterglow_lumi_timing36000}}
\end{figure*}

 Figure \ref{fig:Xray_afterglow_lumi_timing36000} shows the same four plots as of figure~\ref{fig:Xray_afterglow_lumi_timing200} but for the epoch 10 hours after the \textsl{Swift}/BAT trigger time.     The number of the samples is 39.  The correlation coefficients  are $\sigma=0.10_{-0.07}^{+0.08}$, $-0.09_{-0.04}^{+0.05}$, $0.08_{-0.08}^{+0.06}$, and $-0.27_{-0.03}^{+0.05}$, respectively, for the four plots.    In contrast to those at the epoch 200s after the trigger (the previous subsection),  no clear correlations among those properties were found.  Note that we had excluded the data of XRF120724A because  of lack of the data (the available XRT data covered up to only  2.7 hours after the BAT trigger).

\begin{longrotatetable}
\startlongtable
\begin{deluxetable}{l c c c c c c c c c}
\tablecaption{Fitting results of the X-ray afterglow light-curves (0.3--10 keV luminosity).\label{tbl:XrayLumFitResult}}
\tablehead{
  \colhead{Events} & \colhead{$\Gamma_1$\tablenotemark{a}} & \colhead{$t_1$\tablenotemark{b} [sec]} & \colhead{$\Gamma_2$\tablenotemark{c}} & \colhead{$t_2$\tablenotemark{d} [sec]} & \colhead{$\Gamma_3$\tablenotemark{e}} & \colhead{$t_3$\tablenotemark{f} [sec]} & \colhead{$\Gamma_4$\tablenotemark{g}} & \colhead{$\chi^2$/d.o.f.} & \colhead{Best-fit model\tablenotemark{h}}
  }
\startdata
XRF050406 & $ -0.468_{-0.148}^{+0.160} $ & $ 63.3_{-35.9}^{+59.0} $ & $ -0.395_{-0.021}^{+0.022} $ & $ 891_{-85}^{+81} $ & $ -0.871_{-0.014}^{+0.012} $ & \nodata & \nodata & 86.1/94 & brkpow2 \\
XRF050416A & $ -1.55 \pm 0.01 $ & $ 122 \pm 4 $ & $ -0.395_{-0.021}^{+0.022} $ & $ 891_{-85}^{+81} $ & $ -0.871_{-0.014}^{+0.012} $ & \nodata & \nodata & 86.1/94 & brkpow2 \\
XRF050819 & $ -3.25_ \pm 0.02 $ & $ 270_{-58}^{+51} $ & $ -0.344_{-0.055}^{+0.049} $ & $ 8660_{-2740}^{+5600} $ & $ -1.04_{-0.64}^{+0.18} $ & \nodata & \nodata & 30.4/16 & brkpow2 \\
XRF050824 & $ -0.324_{-0.007}^{+0.006} $ & $ 3.53_{-0.49}^{+0.59} \times 10^4  $ & $ -0.869_{-0.054}^{+0.047} $ & \nodata & \nodata & \nodata & \nodata & 48.6/37 & brkpow \\
XRF060926 & $ -1.66_{-0.027}^{+0.024} $ & $ 37.8_{-1.6}^{+2.2} $ & $ 0.056_{-0.015}^{+0.029} $ & $ 300_{-26}^{+20} $ & $ -1.46_{-0.07}^{+0.06} $ & \nodata & \nodata & 6.37/11 & brkpow2 \\
XRF080520 & $ -1.53_{-0.33}^{+0.44} $ & $ 157_{-78}^{+84} $ & $ -0.877_{-0.11}^{+0.092} $ & $ 5140_{-3200}^{+17000} $ & $ -1.28_{-0.55}^{+0.28} $ & \nodata & \nodata & 1.96/3 & brkpow2 \\
XRF081007 & $ -3.73 \pm 0.01 $ & $ 179_{-6}^{+11} $ & $ -0.533_{-0.003}^{+0.03} $ & $ 5630_{-670}^{+630} $ & $ -1.07_{-0.02}^{+0.03} $ & \nodata & \nodata & 84.3/73 & brkpow2 \\
XRF100425A & $ -4.65_{-0.02}^{+0.01} $ & $ 141_{-2}^{+3} $ & $ 0.026_{-0.082}^{+0.077} $ & $ 421_{-54}^{+56} $ & $ -0.651_{-0.032}^{+0.029} $ & $ 2.12_{-0.61}^{+0.58} \times 10^4 $ & $ -1.26_{-0.23}^{+0.16} $ & 17.8/21 & brkpow3 \\
XRF110808A & $ -3.57 \pm 0.01 $ & $ 293_{-9}^{+10} $ & $ -0.106_{-0.041}^{+0.037} $ & $ 5820_{-1000}^{+1100} $ & $ -0.804_{-0.068}^{+0.058} $ & $ (1.29 \pm 0.31) \times 10^5 $ & $ -2.04_{-1.0}^{+0.67} $ & 26.4/28 & brkpow3 \\
XRF120724A & $ -0.471_{-0.127}^{+0.143} $ & $ 29.7_{-16.5}^{+41.8} $ & $ 0.005_{-0.021}^{+0.020} $ & $ 3560_{-140}^{+130} $ & $ -2.84_{-0.27}^{+0.23} $ & \nodata & \nodata & 56.8/34 & brkpow2 \\
XRF130612A & $ -0.467_{-0.284}^{+0.293} $ & $ 152 \pm 23 $ & $ 0.005_{-0.021}^{+0.020} $ & $ 3560_{-140}^{+130} $ & $ -2.84_{-0.27}^{+0.23} $ & \nodata & \nodata & 56.8/34 & brkpow2 \\ \hline
XRR050318 & $ -1.23 \pm 0.01 $ & $ 7330_{-700}^{+670} $ & $ -2.09_{-0.13}^{+0.11} $ & \nodata & \nodata & \nodata & \nodata & 68.4/75 & brkpow \\
XRR050525A & $ -0.711 \pm 0.008 $ & $ 4680_{-450}^{+450} $ & $ -1.55 \pm 0.05 $ & \nodata & \nodata & \nodata & \nodata & 26.8/28 & brkpow \\
XRR060206 & $ -0.925_{-0.015}^{+0.014} $ & $ 320_{-47}^{+52} $ & $ -0.416 \pm 0.031 $ & $ 5920_{-640}^{+620} $ & $ -1.26_{-0.05}^{+0.04} $ & \nodata & \nodata & 12.7/20 & brkpow2 \\
XRR060707 & $ -2.80 \pm 0.01 $ & $ 112 \pm 3 $ & $ -0.535_{-0.024}^{+0.022} $ & $ 7050_{-1500}^{+1700} $ & $ -1.01_{-0.05}^{+0.04} $ & $ 2.46_{-0.36}^{+0.50} \times 10^5 $ & $ -2.35_{-0.69}^{+0.49} $ & 46.6/38 & brkpow3 \\
XRR060927 & $ -0.757_{-0.018}^{+0.017} $ & $ 662_{-110}^{+130} $ & $ -1.53_{-0.32}^{+0.18} $ & \nodata & \nodata & \nodata & \nodata & 8.71/14 & brkpow \\
XRR061222B & $ -3.35 \pm 0.01 $ & $ 95.2_{-6.6}^{+6.1} $ & $ -1.59_{-0.11}^{+0.08} $ & \nodata & \nodata & \nodata & \nodata & 36.3/38 & brkpow \\
XRR071010B & $ -0.663_{-0.063}^{+0.064} $ & \nodata & \nodata  & \nodata & \nodata & \nodata & \nodata & 8.71/14 & PL \\
XRR080207 & $ -1.69 \pm 0.01 $ & $ 4080_{-1400}^{+1900} $ & $ -1.85_{-0.11}^{+0.095} $ & \nodata & \nodata & \nodata & \nodata & 44.1/60 & brkpow \\
XRR080603B & $ -3.50_{-0.01}^{+0.01} $ & $ 40.4_{-0.8}^{+1.0} $ & $ -0.839_{-0.032}^{+0.028} $ & \nodata & \nodata & \nodata & \nodata & 100/92 & brkpow \\
XRR081221 & $ -0.562 \pm 0.004 $ & $ 209 \pm 8 $ & $ -1.28 \pm 0.01 $ & $ 9.16_{-2.49}^{+4.31} \times 10^4 $ & $ -2.32_{-1.30}^{+0.83} $ & \nodata & \nodata & 255/262 & brkpow2 \\
XRR090423 & $ -4.88 \pm 0.02 $ & $ 38.7_{-0.7}^{+0.8} $ & $ -0.366_{-0.028}^{+0.026} $ & $ 1340_{-180}^{+190} $ & $ -1.48_{-0.12}^{+0.08} $ & \nodata & \nodata & 17.5/17 & brkpow2 \\
XRR100615A & $ -4.22 \pm 0.01 $ & $ 81.5 \pm 0.8 $ & $ -0.009 \pm 0.026 $ & $ 454_{-39}^{+45} $ & $ -0.480_{-0.021}^{+0.021} $ & $ 8590_{-840}^{+870} $ & $ -1.22_{-0.05}^{+0.05} $ & 70.7/87 & brkpow3 \\
XRR100621A & $ -3.77 \pm 0.01 $ & $ 274_{-2.3}^{+2.5} $ & $ -0.640_{-0.012}^{+0.011} $ & $ 4050_{-440}^{+370} $ & $ -0.918_{-0.017}^{+0.018} $ & $ 5.72_{-0.50}^{+0.95} \times 10^4 $ & $ -1.52_{-0.06}^{+0.05} $ & 416/356 & brkpow3 \\
XRR110726A & $ -0.853_{-0.051}^{+0.052} $ & \nodata & \nodata & \nodata & \nodata & \nodata & \nodata & 16.7/11 & PL \\
XRR120326A & $ -4.54_{-0.0058}^{+0.0062} $ & $ 93.1_{-0.56}^{+0.56} $ & $ -0.136_{-0.0068}^{+0.0072} $ & $ 4490_{-270}^{+160} $ & $ 0.469_{-0.029}^{+0.025} $ & $ (1.56 \pm 0.03) \times 10^4 $ & $ -1.84_{-0.07}^{+0.06} $ & 176/194 & brkpow3 \\
XRR120802A & $ -2.79 \pm 0.02 $ & $ 53.5_{-1.6}^{+1.8} $ & $ -0.372_{-0.025}^{+0.023} $ & \nodata & \nodata & \nodata & \nodata & 28.9/31 & brkpow \\
XRR120811C & $ -3.22 \pm 0.01 $ & $ 60.0_{-1.1}^{+1.3} $ & $ -0.504_{-0.029}^{+0.028} $ & $ 808 \pm 98 $ & $ -1.19 \pm 0.04 $ & \nodata & \nodata & 140/106 & brkpow2 \\
XRR121128A & $ -4.90 \pm 0.01 $ & $ 43.4_{-0.3}^{+0.4} $ & $ -0.600_{-0.021}^{+0.020} $ & $ 466_{-32}^{+28} $ & $ -1.55_{-0.03}^{+0.04} $ & $ 1.70_{-0.24}^{+0.25} \times 10^4 $ & $ -3.90_{-1.80}^{+1.21} $ & 117/119 & brkpow3 \\
XRR130701A & $ -2.07 \pm 0.01 $ & $ 61.8_{-1.3}^{+1.4} $ & $ -0.703 \pm 0.040 $ & $ 195_{-18}^{+17} $ & $ -1.27 \pm 0.02 $ & \nodata & \nodata & 119/124 & brkpow2 \\ 
XRR130925A & $ -2.43 \pm 0.02 $ & $ 903_{-12}^{+11} $ & $ -0.832 \pm 0.004 $ & $ 3.11_{-0.19}^{+0.25} \times 10^5 $ & $ -1.30_{-0.03}^{+0.02} $ & \nodata   & \nodata   & 900/799 & brkpow2 \\  \hline
GRB080804 & $ -1.09 \pm 0.01 $ & $ 1.71_{-0.87}^{+2.10} \times 10^4 $ & $ -1.32_{-0.20}^{+0.15} $ & \nodata & \nodata & \nodata & \nodata & 77.1/99 & brkpow \\
GRB080916A & $ -3.31 \pm 0.01 $ & $ 172 \pm 3 $ & $ -0.856_{-0.013}^{+0.016} $ & $ 2.42_{-0.51}^{+0.78} \times 10^4 $ & $ -1.18_{-0.05}^{+0.06} $ & \nodata & \nodata & 175/130 & brkpow2 \\
GRB081121 & $ -1.43 \pm 0.02 $ & \nodata & \nodata & \nodata & \nodata & \nodata & \nodata & 193/145 & PL \\
GRB081222 & $ -2.14 \pm 0.01 $ & $ 20.9 \pm 0.2 $ & $ -0.837 \pm 0.001 $ & $ 148_{-10}^{+11} $ & $ -1.11 \pm 0.01 $ & $ 2.08_{-0.29}^{+0.30} \times 10^4 $ & $ -1.97_{-0.14}^{+0.12} $ & 455/416 & brkpow3 \\
GRB090102 & $ -0.977 \pm 0.005 $ & $ 575_{-48}^{+51} $ & $ -1.45_{-0.02}^{+0.01} $ & \nodata & \nodata & \nodata & \nodata & 139/138 & brkpow \\
GRB090424 & $ -2.20 \pm 0.01 $ & $ 97.9 \pm 1.0 $ & $ -0.800_{-0.008}^{+0.009} $ & $ 920_{-62}^{+53} $ & $ -1.18 \pm 0.01 $ & \nodata & \nodata & 542/663 & brkpow2 \\
GRB090926B & $ -2.27_{-0.13}^{+0.11} $ & $ 290_{-44}^{+54} $ & $ -1.09_{-0.13}^{+0.11} $ & \nodata &  \nodata & \nodata & \nodata & 75.4/88 & brkpow \\
GRB100816A & $ -1.03 \pm 0.04 $ & \nodata & \nodata & \nodata & \nodata & \nodata & \nodata & 24.2/27 & PL \\
GRB110731A & $ -1.16 \pm 0.01 $ & $ 1860_{-600}^{+1200} $ & $ -1.29 \pm 0.04 $ & \nodata & \nodata & \nodata & \nodata & 261/249 & brkpow \\
\enddata
\end{deluxetable}
\tablenotetext{a}{Decay index of the 1st power-law component.}
\tablenotetext{b}{Break time of the 1st component in seconds after the BAT trigger.}
\tablenotetext{c}{Decay index of the 2nd power-law component.}
\tablenotetext{d}{Break time of the 2nd component in seconds after the BAT trigger.}
\tablenotetext{e}{Decay index of the 3rd power-law component.}
\tablenotetext{f}{Break time of the 3rd component in seconds after the BAT trigger.}
\tablenotetext{g}{Decay index of the 4th power-law component.}
\tablenotetext{h}{Best-fit  model. The models ``brkpow'',  ``brkpow2'', and ``brkpow3''  have two, three, and four decay  indices of power-law components, respectively.}
\end{longrotatetable}

\begin{longrotatetable}
\startlongtable
\begin{deluxetable}{l c c c c c c}
\tablecaption{ X-ray luminosity and  temporal decay index  200, 3600, and 36000 sec after the BAT trigger time in the rest frame of the GRBs.\label{tbl:XrayLumResult2}}
\tablehead{
  \colhead{Events} & \colhead{$L_{200}$\tablenotemark{a}} & \colhead{$\Gamma_{200}$\tablenotemark{b}} & \colhead{$L_{3600}$\tablenotemark{c}} & \colhead{$\Gamma_{3600}$\tablenotemark{d}} & \colhead{$L_{36000}$\tablenotemark{e}} & \colhead{$\Gamma_{36000}$\tablenotemark{f}}
}
\startdata
XRF050406 & $ 63.3_{-35.9}^{+59.0} $ & $ -0.468_{-0.148}^{+0.160} $ & $ 8.61_{-7.53}^{+33.00} $ & $ -0.468_{-0.148}^{+0.160} $ & $ 2.93_{-2.55}^{+17.40} $ & $ -0.468_{-0.148}^{+0.160} $   \\ 
XRF050416A & $ 52.0_{-29}^{+88} $ & $ -0.395 \pm 0.021 $ & $ 8.54_{-0.69}^{+7.40} $ & $ -0.871_{-0.014}^{+0.012} $ & $ 1.15_{-0.06}^{+1.1} $ & $ -0.871_{-0.014}^{+0.012} $ \\ 
XRF050819 & $ 48.8_{-46}^{+14} $ & $ -0.344_{-0.069}^{+0.042} $ & $ 18.1_{-2.9}^{+2.9} $ & $ -0.344_{-0.056}^{+0.048} $ & $ 3.64_{-2.40}^{+1.30} $ & $ -1.01_{-0.65}^{+0.16} $ \\
XRF050824 & $ 13.8_{-4.5}^{+49} $ & $ -0.324_{-0.069}^{+0.069} $ & $ 4.77_{-1.10}^{+19.00} $ & $ -0.324_{-0.006}^{+0.007} $ & $ 2.24_{-2.20}^{+2.20} $ & $ -0.869_{-0.054}^{+0.047} $ \\ 
XRF060926 & $ 3800_{-680}^{+680} $ & $ 0.056_{-0.015}^{+0.029} $ & $ 104_{-12}^{+12} $ & $ -1.46_{-0.26}^{+0.20} $ & $ 3.63 \pm 0.44 $ & $ -1.46_{-0.26}^{+0.20} $ \\ 
XRF080520 & $ 218_{-130}^{+360} $ & $ -0.877_{-0.110}^{+0.09} $ & $ 17.3_{-13.0}^{+39.0} $ & $ -0.877_{-0.110}^{+0.092} $ & $ 1.37_{-1.10}^{+4.00} $ & $ -1.28_{-0.55}^{+0.28} $ \\ 
XRF081007 & $ 44.3_{-21.0}^{+56.0} $ & $ -0.519_{-0.018}^{+0.017} $ & $ 9.88_{-5.60}^{+6.30} $ & $ -0.533_{-0.003}^{+0.032} $ & $ 1.09_{-0.35}^{+2.20} $ & $ -1.07_{-0.02}^{+0.03} $ \\ 
XRF100425A & $ 100_{-11}^{+32} $ & $ 0.026_{-0.030}^{+0.077} $ & $ 25.4_{-18}^{+64} $ & $ -0.651_{-0.032}^{+0.029} $ & $ 4.12_{-0.82}^{+0.69} $ & $ -1.26_{-0.23}^{+0.16} $ \\ 
XRF110808A & $ 20.2_{-5.2}^{+3.0} $ & $ -0.106_{-0.034}^{+0.030} $ & $ 14.8_{-3.2}^{+27.0} $ & $ -0.106_{-0.041}^{+0.037} $ & $ 3.26_{-2.7}^{+3.00} $ & $ -0.804_{-0.068}^{+0.058} $ \\
XRF120724A & $ 29.7_{-16.5}^{+41.8} $ & $ -0.471_{-0.127}^{+0.143} $ & $ 7.62_{-5.27}^{+20.1} $ & $ -0.155_{-0.080}^{+0.068} $ & \nodata & \nodata \\ 
XRF130612A & $ 153 \pm 23 $ & $ -0.467_{-0.284}^{+0.293} $ & $ 16.2 \pm 2.4 $ & $ -1.22_{-0.22}^{+0.20} $ & $ 0.986 \pm 0.148 $ & $ -1.22_{-0.22}^{+0.20} $  \\ \hline
XRR050318 & \nodata & \nodata & $ 135_{-38}^{+48} $ & $ -1.23 \pm 0.04 $ & $ 2.05_{-0.37}^{+0.40} $ & $ -2.09_{-0.13}^{+0.11} $ \\ 
XRR050525A & $ 234_{-78}^{+97} $ & $ -0.711 \pm 0.008 $ & $ 30.0_{-10.1}^{+12.2} $ & $ -0.711_{-0.0081}^{+0.0076} $ & $ 1.05_{-0.52}^{+1.2} $ & $ -1.55 \pm 0.053 $ \\ 
XRR060206 & $ 4560_{-810}^{+870} $ & $ -0.416_{-0.043}^{+0.051} $ & $ 1340_{-530}^{+100} $ & $ -0.416 \pm 0.031 $ & $ 112_{-69}^{+210} $ & $ -1.26_{-0.05}^{+0.04} $ \\ 
XRR060707 & $ 2130_{-1000}^{+1800} $ & $ -0.535_{-0.033}^{+0.029} $ & $ 452_{-210}^{+380} $ & $ -0.535_{-0.024}^{+0.022} $ & $ 60.7_{-40}^{+140} $ & $ -1.01_{-0.05}^{+0.04} $ \\ 
XRR060927 & $ 6630_{-2200}^{+9200} $ & $ -0.758_{-0.070}^{+0.170} $ & $ 200_{-83}^{+70} $ & $ -1.53_{-0.32}^{+0.18} $ & $ 5.88_{-4.20}^{+6.22} $ & $ -1.53_{-0.32}^{+0.18} $ \\ 
XRR061222B & $ 1670_{-750}^{+1600} $ & $ -1.59_{-0.11}^{+0.08} $ & $ 16.7_{-7.5}^{+16.0} $ & $ -1.59_{-0.11}^{+0.08} $ & $ 0.426_{-0.191}^{+0.412} $ & $ -1.59_{-0.11}^{+0.08} $ \\ 
XRR071010B & \nodata & \nodata & $ 50.3_{-31.2}^{+54.1} $ & $ -0.663_{-0.063}^{+0.064} $ & $ 10.9_{-7.4}^{+13.8} $ & $ -0.663_{-0.063}^{+0.064} $ \\ 
XRR080207 & $ 37600_{-28000}^{+25000} $ & $ -1.78_{-0.07}^{+0.10} $ & $ 287_{-190}^{+460} $ & $ -1.69 \pm 0.006 $ & $ 4.15_{-3.00}^{+10.25} $ & $ -1.85_{-0.11}^{+0.1-} $ \\ 
XRR080603B & $ 7960_{-830}^{+2100} $ & $ -0.835 \pm 0.041 $ & $ 704_{-110}^{+140} $ & $ -0.839 \pm 0.03 $ & $ 102_{-16}^{+20} $ & $ -0.839_{-0.032}^{+0.028} $ \\ 
XRR081221 & $ 15600_{-6900}^{+7100} $ & $ -0.562 \pm 0.004 $ & $ 392_{-61}^{+43} $ & $ -1.28 \pm 0.01 $ & $ 20.3_{-3.2}^{+2.3} $ & $ -1.28 \pm 0.01 $ \\ 
XRR090423 & $ 9360_{-3100}^{+1100} $ & $ -0.366_{-0.029}^{+0.026} $ & $ 690_{-490}^{+1400} $ & $ -1.42 \pm 0.06 $ & $ 26.1_{-21.1}^{+67.0} $ & $ -1.42 \pm 0.06 $ \\ 
XRR100615A & $ 1360_{-90}^{+310} $ & $ -0.00871_{-0.02690}^{+0.00831} $ & $ 498_{-270}^{+590} $ & $ -0.480 \pm 0.0210 $ & $ 57.2_{-43.2}^{+200.5} $ & $ -1.22 \pm 0.05 $ \\ 
XRR100621A & $ 340_{-240}^{+85} $ & $ -0.640 \pm 0.140 $ & $ 53.4_{-38.0}^{+13.2} $ & $ -0.640_{-0.012}^{+0.011} $ & $ 6.67_{-3.40}^{+2.80} $ & $ -0.918_{-0.017}^{+0.018} $ \\ 
XRR110726A & $ 63.3_{-27.4}^{+38.0} $ & $ -0.853_{-0.051}^{+0.052} $ & $ 5.38_{-2.67}^{+3.96} $ & $ -0.853_{-0.051}^{+0.052} $ & $ 0.755_{-0.417}^{+0.670} $ & $ -0.853_{-0.051}^{+0.052} $  \\ 
XRR120326A & $ 263_{-160}^{+450} $ & $ -0.139_{-0.013}^{+0.012} $ & $ 149_{-110}^{+320} $ & $ -0.136 \pm -0.007 $ & $ 55.8_{-28.0}^{+67.0} $ & $ -1.84_{-0.067}^{+0.059} $ \\ 
XRR120802A & $ 1270_{-430}^{+570} $ & $ -0.372_{-0.055}^{+0.058} $ & $ 432_{-150}^{+200} $ & $ -0.372_{-0.025}^{+0.023} $ & $ 183_{-62}^{+83} $ & $ -0.372_{-0.025}^{+0.023} $ \\ 
XRR120811C & $ 6420_{-2600}^{+8000} $ & $ -0.474_{-0.057}^{+0.054} $ & $ 537_{-380}^{+1100} $ & $ -1.19 \pm 0.04 $ & $ 34.7_{-27.0}^{+110.0} $ & $ -1.19 \pm 0.043 $ \\ 
XRR121128A & $ 6740_{-1600}^{+2200} $ & $ -0.600_{-0.022}^{+0.038} $ & $ 171_{-62}^{+70} $ & $ -1.55_{-0.04}^{+0.03} $ & $ 0.842_{-0.620}^{+1.211} $ & $ -3.90_{-1.80}^{+1.22} $ \\ 
XRR130701A & $ 3120_{-960}^{+1400} $ & $ -1.27_{-0.05}^{+0.04} $ & $ 80.4_{-24.0}^{+35.0} $ & $ -1.27 \pm 0.02 $ & $ 4.36_{-1.31}^{+1.92} $ & $ -1.27 \pm 0.02 $ \\ 
XRR130925A & $ 631_{-190}^{+240} $ & $ -0.828 \pm 0.004 $ & $ 57.6_{-17.3}^{+21.2} $ & $ -0.828_{-0.0035}^{+0.0037} $ & $ 8.56_{-2.60}^{+3.20} $ & $ -0.828 \pm 0.004 $ \\ \hline
GRB080804 & $ 2410_{-89}^{+88} $ & $ -1.09 \pm 0.01 $ & $ 103_{-11}^{+12} $ & $ -1.09 \pm 0.01 $ & $ 7.12 \pm 1.1 $ & $ -1.32_{-0.20}^{+0.15} $ \\ 
GRB080916A & $ 158_{-52}^{+90} $ & $ -0.856_{-0.013}^{+0.016} $ & $ 13.4_{-3.8}^{+9.3} $ & $ -0.856_{-0.013}^{+0.016} $ & $ 1.64_{-0.71}^{+4.71} $ & $ -1.18_{-0.05}^{+0.06} $ \\ 
GRB081121 & $ 80400_{-14000}^{+17000} $ & $ -1.43 \pm 0.02 $ & $ 1290_{-230}^{+280} $ & $ -1.43 \pm 0.02 $ & $ 47.6_{-8.4}^{+10.0} $ & $ -1.43 \pm 0.02 $ \\ 
GRB081222 & $ 15900_{-1400}^{+2100} $ & $ -1.11 \pm 0.02 $ & $ 645_{-59}^{+81} $ & $ -1.11 \pm 0.01$ & $ 31.5_{-4.2}^{+3.5} $ & $ -1.97_{-0.14}^{+0.12} $ \\ 
GRB090102 & $ 8340_{-4700}^{+11000} $ & $ -0.977_{-0.016}^{+0.017} $ & $ 208_{-58}^{+81} $ & $ -1.45_{-0.02}^{+0.01} $ & $ 7.43_{-2.10}^{+2.90} $ & $ -1.45_{-0.02}^{+0.01} $ \\ 
GRB090424 & $ 1800_{-260}^{+290} $ & $ -0.800 \pm 0.01 $ & $ 106_{-10}^{+12} $ & $ -1.18 \pm 0.01 $ & $ 7.10_{-0.64}^{+0.83} $ & $ -1.18 \pm 0.01 $ \\ 
GRB090926B & $ 740_{-286}^{+435} $ & $ -1.09_{-0.10}^{+0.11} $ & $ 33.6_{-19.6}^{+40.5} $ & $ -1.09_{-0.10}^{+0.11} $ & $ 2.71_{-1.75}^{+4.56} $ & $ -1.09_{-0.10}^{+0.11} $ \\ 
GRB100816A & $ 149_{-6}^{+110} $ & $ -1.04_{-0.03}^{+0.05} $ & $ 6.14_{-1.80}^{+3.60} $ & $ -1.03 \pm 0.04 $ & $ 0.572_{-0.180}^{+0.400} $ & $ -1.03 \pm 0.04 $ \\ 
GRB110731A & $ 12100_{-800}^{+2200} $ & $ -1.15_{-0.02}^{+0.03} $ & $ 395_{-140}^{+610} $ & $ -1.29 \pm 0.041 $ & $ 20.2_{-7.1}^{+31.1} $ & $ -1.29 \pm 0.04 $ \\
\enddata
\tablenotetext{a}{0.3--10 keV luminosity ($10^{45}$ erg s$^{-1}$)  200 s after the BAT trigger time. }
\tablenotetext{b}{Decay index in the afterglow light-curve of 0.3--10 keV luminosity  200 sec after the BAT trigger time.}
\tablenotetext{c}{0.3--10 keV luminosity ($10^{45}$ erg s$^{-1}$)  3600 s after the BAT trigger time. }
\tablenotetext{d}{Decay index in the afterglow light-curve of 0.3--10 keV luminosity  3600 sec after the BAT trigger time.}
\tablenotetext{e}{0.3--10 keV luminosity ($10^{45}$ erg s$^{-1}$)  36000 s after the BAT trigger time.}
\tablenotetext{f}{Decay index in the afterglow light-curve of 0.3--10 keV luminosity  36000 sec after the BAT trigger time. }
\end{deluxetable}
\end{longrotatetable}

\section{DISCUSSION} \label{sec:discussion}
\subsection{Comparison of the observed afterglow light-curves with the boxfit simulation}
  We used the  ``boxfit'' \citep{2012ApJ...749...44V} tool to perform simulations of the afterglows for their light curves in order to verify whether the origin of the $E_{{\rm peak}}$ diversity in prompt emission is the properties of jet or the geometrical effect. 
  \subsubsection{The variable opening angle jet model}
\begin{figure}[h]
  \centering
  \includegraphics[width=6cm,angle=270]{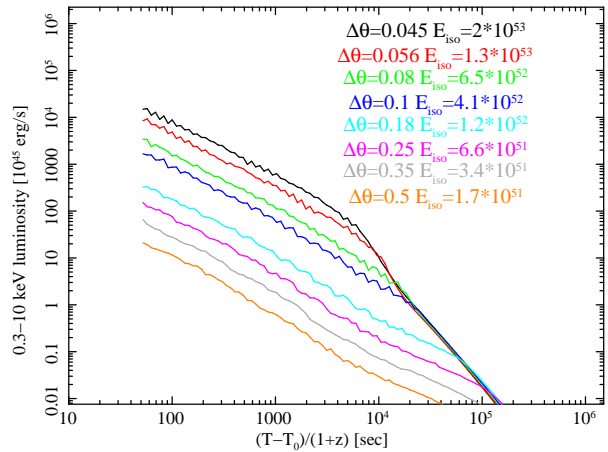}
\caption{Light curves simulated  with the boxfit tools on the basis of the variable opening-angle model, where  the jet opening angle $\Delta \theta$ and $E_{{\rm iso}}$ are allowed to vary. The fixed parameters  are $\theta_{{\rm obs}}=0$, $n=1$, $p=2.5$, $\epsilon_B=10^{-5}$, and $\epsilon_e=0.2$.\label{fig:sim_opening_angle_mod}}
\end{figure}

  The variable opening angle jet model is one of the model which explains the $E_{{\rm peak}}$ diversity by the properties of jet itself. Figure \ref{fig:sim_opening_angle_mod} shows the simulated light curves of the X-ray afterglow   with the tools on the basis of the variable opening-angle model, where  the jet opening angles $\Delta \theta$ are allowed to vary.   In the simulations, we have used the values of  $E_{{\rm iso}}$  calculated from following equation given by  \citet{2005ApJ...620..355L},
   \begin{equation} \label{variable_eiso}
       E_{{\rm iso}} = \frac{E_{\gamma}}{1-\cos\Delta\theta},
   \end{equation}
   where $E_{\gamma}$ is the energy of emitted photon energy and we assumed it $E_{\gamma}=1.3\times10^{51}$, estimated by \citet{2003ApJ...588..945B}.  The  events with a larger $\Delta \theta$ (or lower $E_{{\rm iso}}$)  are found to  have a lower X-ray luminosity than those with a  smaller $\Delta \theta$ (higher $E_{{\rm iso}}$).   Also, figure  \ref{fig:sim_opening_angle_mod} implies that the break time of the jets  with a smaller $\Delta \theta$ comes earlier.  The positive correlation between the X-ray luminosity and $E_{{\rm iso}}$ (or ``hardness'' of the events)  exists only in the early  phase of afterglows and disappears in the  later phase.  Accordingly, the simulated X-ray luminosity light-curves  on the basis of  the variable opening-angle model are consistent with the observed $L_{{\rm X}}$-$E_{{\rm peak}}^{{\rm src}}$ ($E_{{\rm iso}}$) relations.
   
   The range of $E_{{\rm iso}}$ in our sample, however, is $10^{50}<E_{{\rm iso}}<10^{54}$ erg, and the lower limit is 10 times smaller than $E_{\gamma}$ estimated by \citet{2003ApJ...588..945B}.  Thus, there is no value of the opening angle that can accommodate this value of $E_{{\rm iso}}$. According to \citet{2005ApJ...620..355L}, in order to escape this problem, $E_{\gamma}$ is rescaled and the equation \ref{variable_eiso} is modified to be
  \begin{equation} \label{variable_eiso_2}
       E_{{\rm iso}} = \frac{E_{\gamma}}{95\times(1-\cos\Delta\theta)} .
  \end{equation}
  In the basis of this equation, we obtained the range of $\Delta\theta$, corresponding to that of $E_{{\rm iso}}$. The range is $5.2 \times 10^{-3}<\Delta\theta<0.53$ rad.
  
  Here, we estimate range of jet break time ($t_{{\rm jet}}$) from those of $\Delta \theta$ and of $E_{{\rm iso}}$. $t_{{\rm jet}}$ means the break time in the light curve of the afterglow. After the $t_{{\rm jet}}$, the temporal decay index become smaller than -2 \citep{1999ApJ...519L..17S}.   The $t_{{\rm jet}}$ is given by the equation (1) in \citet{1999ApJ...519L..17S},
\begin{equation} \label{eq:jet_break_time}
    t_{{\rm jet}} \approx 6.2(E_{{\rm iso},52}/n)^{1/3}(\Delta\theta/0.1)^{8/3},
\end{equation}
where $E_{{\rm iso},52}$ is the isotropic equivalent energy in units of $10^{52}$ ergs. If we assume circum-burst number density $n=1$ [cm$^{-3}$], the range of $t_{{\rm jet}}$ is $40 <t_{{\rm jet}}<4 \times 10^5\, {\rm sec}$. Therefore, It is necessary that jet breaks are occurred in afterglow light curves of energetic GRBs ($E_{{\rm iso}} > 2.0 \times 10^{54}$ erg) before 200 sec after trigger. The results shown in \S\ref{ss:xrt_Xray_afterglow_result}, however, are inconsistent with the prediction of such a early jet break.
  
\subsubsection{The off-axis jet model}
\begin{figure*}
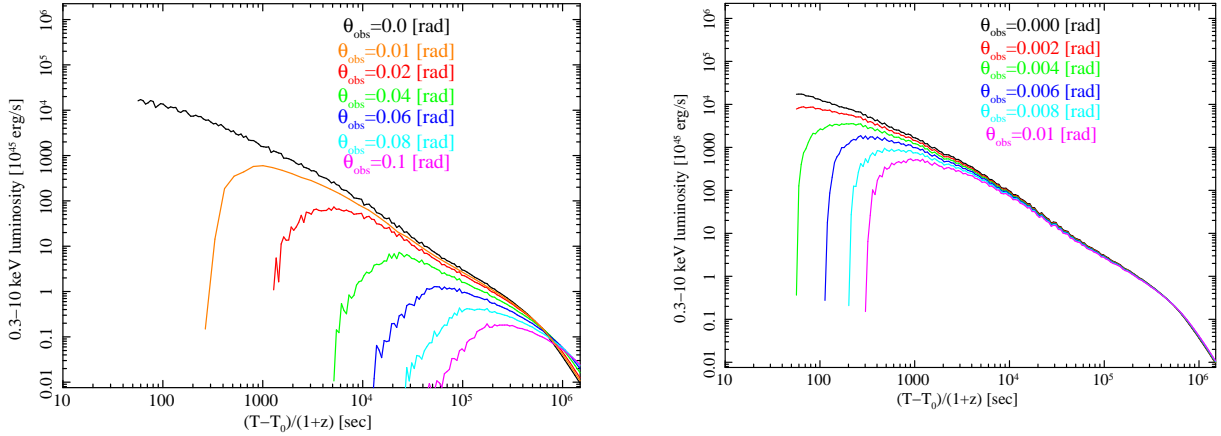

\gridline{
\includegraphics[width=5.8cm,angle=270]{xrf081007_sim_off_axis_x_large_lum.eps}
\includegraphics[width=5.5cm,angle=270]{xrf081007_sim_off_axis_x_lum.eps}
}
\caption{ Simulated light curves  with the boxfit tools on the basis of the off-axis model with the varying observing angle $\theta_{{\rm obs}}$ for a range of (\textsl{left})  0--0.1 rad ($\approx6^\circ$),  drawn every 0.02 rad, (\textsl{right})  0--0.01 rad ($\approx0.6^\circ$),  drawn every 0.002 rad. See figure~\ref{fig:sim_opening_angle_mod} caption for the fixed parameters. \label{fig:sim_off_axis_mod}}
\end{figure*}

Another likely model of GRBs is the off-axis model, in which the situation of an observer being off-axis from the jet is considered. The off-axis model predicts an existence of a rising part in the afterglow light curve because its beamed emission is less visible to the observer  in the early  phase. Figure \ref{fig:sim_off_axis_mod} shows the X-ray luminosity light-curves simulated  with the boxfit tools on the basis of the off-axis model. 
 The results suggest the trend that  the peak of light curves and  start time of the rising part  come later for the events with a wider $\theta_{{\rm obs}}$.  The trend is consistent with the results from our observed samples (see \S\ref{ss:xrt_Xray_afterglow_result}), which have indicated   that the X-ray luminosity of  the XRF samples  are  lower than  those of C-GRBs. However,  our observation results  (figures 8--10)  do not show any significant rising parts  as in figure \ref{fig:sim_off_axis_mod}(a). Therefore, these results  suggest that the diversity  in the observing angles  has to be restricted in  a very narrow range of $\theta_{{\rm obs}} < 0.01$ rad ($\approx0.6^\circ$) on the basis of the off-axis model.
\label{ss:discussion_boxfit}

\begin{figure}
\plotone{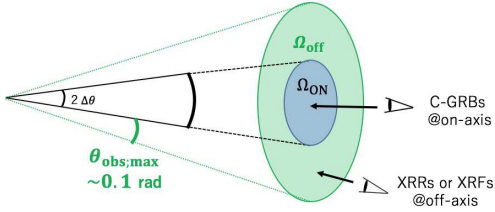}
\caption{ Assumed geometry of the GRB jet in  \S\ref{sec:discussion}. We assume that C-GRBs are observed in the on-axis (blue) area,  whereas XRFs and XRRs are observed in the off-axis (green) area. The parameter $\Omega_{{\rm ON}}$ ($\Omega_{{\rm OFF}}$) is the solid angle subtended by the direction to which  a source is
observed as  a C-GRB (XRR or XRF), and $\theta_{{\rm obs;max}}$ is the observing angle at which hard GRBs ($E_{{\rm peak}}^{{\rm obs}}=100$ keV) are observed as XRFs ($E_{{\rm peak}}^{{\rm obs}}=1$). \label{fig:jet_setup}}
\end{figure}

\subsection{Total number of XRFs, XRRs, and C-GRBs in the whole universe}
To restrict the viable theoretical models, we estimate the total numbers of the XRFs, XRRs, and C-GRBs in the whole universe per year, using the simulator publicly available by \citet{2016ApJ...818...55G},  based on the \textsl{Swift}/BAT trigger algorithm  \citep[][]{2014ApJ...783..24L, 2016ApJ...818...55G}. \rm We execute the simulator setting the maximum likelihood parameters as given in \citep{2016ApJ...818...55G} (the result of the random forest, figure 15). In consequence, we obtain the total numbers of C-GRBs $f_{{\rm C-GRB}}=570 \pm 36$ [events yr$^{-1}$], XRRs $f_{{\rm XRR}}=3031 \pm 53$ [events yr$^{-1}$], and XRFs $f_{{\rm XRF}}=968 \pm 45$ [events yr$^{-1}$], where the  errors  are determined from the Gaussian distribution  obtained after running the simulator 10000 times.

Here, the intrinsic local GRB event rate which we used is $n_0 = 0.42$ [events Gpc$^{-3}$ yr$^{-1}$], derived in \citep{2016ApJ...818...55G}. This rate is consistent with the rate of ``high-luminosity GRBs'', corresponding to C-GRBs and XRRs, of $\sim0.45$ [events Gpc$^{-3}$ yr$^{-1}$]. In contrast, the rate of sub-energetic GRBs (GRB980425 and GRB060218) of $230_{-190}^{+490}$ [events Gpc$^{-3}$ yr$^{-1}$] \citep{2006Natur.442.1014S} and the rate of $10.6$ [events Gpc$^{-3}$ yr$^{-1}$] as measured by \textsl{HETE}-2 \citep{2008A&A...491..157P} are $\sim1000$ and $\sim25$ times higher, respectively, than the rate which we used for the estimation. Especially in the latter one, the population of gamma-ray bursts is dominated by the X-ray flashes. Thus, the estimated total number of the XRFs is a lower limit because the previous studies suggest that the total number of the XRFs was underestimated.
\par
Next, we calculate the canonical opening angle  $\Delta \theta$ and bulk Lorentz factor $\Gamma$ of the jets  from the obtained total numbers. 
We assume that the on-axis GRBs are observed as C-GRBs in an area of $\Omega_{{\rm on}}$ and off-axis ones as XRRs or XRFs in an area of $\Omega_{{\rm off}}$,  as illustrated in figure \ref{fig:jet_setup}. Their ratio is given by, according to \citet{2002ApJ...571L..31Y},
\begin{equation}\label{eq:omega_theta}
 \frac{\Omega_{{\rm off}}}{\Omega_{{\rm on}}} = \frac{2\pi[1-\cos(\theta_{{\rm obs;max}}+\Delta \theta)] - 2\pi[1-\cos (\Delta \theta)]}{2\pi[1-\cos(\Delta \theta)]},
\end{equation}
where  $\theta_{{\rm obs;max}}$ is the observing angle at which $E_{{\rm peak}}$ is observed as 1 keV due to the relativistic Doppler effect. 
We consider that the C-GRBs, XRRs, and XRFs have $E_{{\rm peak}}^{{\rm obs}} > 100$ keV,  $100 \ge E_{{\rm peak}}^{{\rm obs}} > 30$ keV, and  $30 \ge E_{{\rm peak}}^{{\rm obs}} \ge 1$ keV, respectively, on the basis of the results of \S\ref{ss:prompt_results}.  According to \S4 in \citet{2002ApJ...571L..31Y}, the quantity $f_{{\rm XRF}} + f_{{\rm XRR}}$ ($f_{{\rm C-GRB}}$) is the solid angle subtended by the direction to which the source is observed as  an XRF  or XRR (C-GRB).  Thus, the ratio of the solid angles  of  an off-axis to on-axis observers can be described as the ratio of the numbers of the off-axis  to on-axis events,
\begin{equation}
\frac{\Omega_{{\rm off}}}{\Omega_{{\rm on}}} = \frac{f_{{\rm XRR}}+f_{{\rm XRF}}}{f_{{\rm C-GRB}}} \gtrsim 7. 
\end{equation}
We  substitute $\theta_{{\rm obs;max}} \sim 0.01$, which  is estimated in \S\ref{ss:discussion_boxfit}, into  equation (\ref{eq:omega_theta}), 
 solve it  for $\Delta \theta$, and find that
\begin{equation}
  \Delta \theta \lesssim 5.5 \times 10^{-3} [{\rm rad}] \approx 0.32^\circ.
\end{equation}
The $f_{{\rm C-GRB}}$ is identified with the total number of the jets pointed to the Earth per year, which are launched from the sources in the whole universe, such as core-collapse supernovae. Thus, $f_{{\rm src}}=f_{{\rm C-GRB}}/\Omega_{{\rm on}}\approx6.1\times10^6$ is corresponding to the total number of the jets which are launched from the source in the whole universe per year.

Here, we estimate jet break time ($t_{{\rm jet}}$) of XRF from the $\Delta \theta$, using equation \ref{eq:jet_break_time}. If we assume circum-burst number density $n=1$ [cm$^{-3}$] and the energy of the jet as $E_{{\rm iso}}=4.4 \times 10^{53}$ [erg]\footnote{This value is the maximum in our sample (GRB110731A).}, the jet break time is $t_{{\rm jet}} \lesssim 30$ sec.  However, none of XRFs in our sample showed the significant feature of jet break (see in \S\ref{ss:xrt_Xray_afterglow_result}).

The relation between $E_{{\rm peak}}$ of  an on-axis ($E_{{\rm peak}}^{{\rm on}}$) and  off-axis ($E_{{\rm peak}}^{{\rm off}}$) observers is given by \citep{2002ApJ...571L..31Y},
\begin{equation}  \label{eq:off_axis_ep}
  E_{{\rm peak}}^{{\rm off}} = \frac{\delta(1-\beta)}{\Gamma} E_{{\rm peak}}^{{\rm on}}=\frac{1-\beta}{1-\beta\cos(\theta_{{\rm obs}})}E_{{\rm peak}}^{{\rm on}}
\end{equation}
 where $\beta$ is the velocity of the jet and $\delta$ is the Doppler factor.  Substituting $E_{{\rm peak}}^{{\rm off}}=1$ keV, $E_{{\rm peak}}^{{\rm on}}=100$ keV, and $\theta_{{\rm obs}}=\theta_{{\rm obs;max}}\sim 0.01$ into equation \ref{eq:off_axis_ep}, we find the bulk Lorentz factor $\Gamma\approx 1000$.
 
 If we assume to be $E_{{\rm peak}}^{{\rm off}}=10$ keV, the total number of the XRFs and the ratio of solid angle decrease to $f_{{\rm XRF;E_{peak}>10keV}}=539 \pm 36$ [events yr$^{-1}$] and $\Omega_{{\rm off}} / \Omega_{{\rm on}}\gtrsim6$, respectively. The values of $\Delta\theta$, $t_{{\rm jet}}$, and $\Gamma$, corresponding to be based on this condition are $\Delta\theta\lesssim 5.9\times10^{-3} \,{\rm [rad]}\approx 0.34^\circ$, $t_{{\rm jet}}\lesssim 40\, {\rm s}$, and $\Gamma \approx 300$. This $\Gamma$ is smaller than the value derived in the basis of the assumption to be $E_{{\rm peak}}^{{\rm off}}=1$ keV. On the other hand, the results that $\Delta \theta$ is narrow and $t_{{\rm jet}}$ is too fast, are same as the former one. 
 
 In summary, a XRF  is observed when a narrow ($\Delta \theta \sim 0.3^\circ$) jet is  viewed at $\theta_{{\rm obs}} \sim 0.6^{\circ}$.  Thus, the $E_{{\rm peak}}$ diversity, which  is apparent in the BAT samples, needs to be explained  despite a very small variation  in the jet viewing angle, $0<\theta_{{\rm obs}} \lesssim 0.6^{\circ}$, of the XRFs.  This result supports the conclusion by  \citet{2006ApJ...645..436D}, in which the mechanism generating the $E_{{\rm peak}}^{{\rm src}}-E_{{\rm iso}}$ relation \citep{2002A&A...390..81A} was discussed. Additionally, the $t_{{\rm jet}}$ corresponding to estimated $\Delta \theta$ is 30s (or 40s) and none of XRFs in our sample showed the significant feature of jet break. Therefore, the $E_{{\rm peak}}$ diversity among GRBs observed by \textsl{Swift}/BAT, \textsl{Fermi}/GBM and others  are likely to mainly originate 
  not from the off-axis effect but rather the properties of the jet itself, e.g., the variable opening angles. 

\label{ss:discussion_number}

\subsection{Multi-band light curve fitting with the boxfit tools.}
  In order to not only constrain values of $\Delta\theta$ and $\theta_{obs}$ more strongly than the estimations from the event rates but also to find out whether the optical and X-ray afterglow are the same component of external shock models (as expected in the standard model e.g., \citet{2000ApJ...537L..191F}), we executed a model fit for the boxfit-simulated  data  for the X-ray and optical afterglows  originally observed by \textsl{Swift}/XRT and optical telescopes on the ground, respectively. Table~\ref{tbl:opt_sample} summarizes the samples used.  Those samples were selected from our analyzed samples whose optical data were rich.


 For the data of the C-GRBs, we used the parameter region of $E_{{\rm iso}}$ derived from our analysis of the prompt emissions.  For the other  data, while the lower limit of $E_{{\rm iso}}$  was set the same  as those for the C-GRBs, we did not set the upper limit because the simulated jet energy $E_j$ in the boxfit tools relates to  $E_{{\rm iso}}$  by  
\begin{equation}\label{eq:box_jet_ene}
E_j = E_{{\rm iso}}(1-\cos\Delta\theta)\approx E_{{\rm iso}}\Delta\theta^2/2,
\end{equation}
 and because we have to consider the attenuation of $E_{{\rm iso}}$ caused by off-axis effects in order not to underestimate the jet energy. 

\begin{figure*}
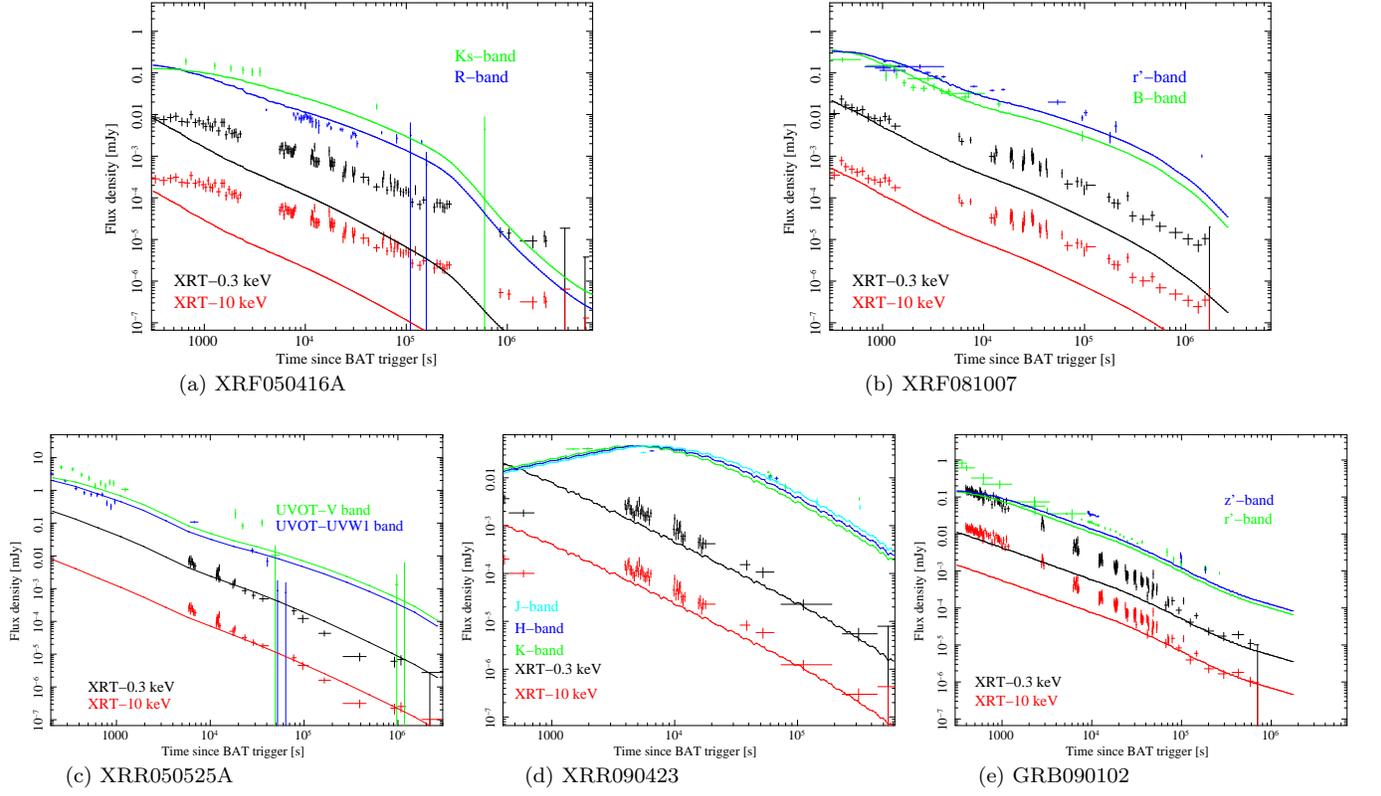

\gridline{\rotatefig{270}{XRF050416A_monoflux_allband.eps}{0.27\textwidth}{(a) XRF050416A}
  \rotatefig{270}{XRF081007_monoflux_allband.eps}{0.27\textwidth}{(b) XRF081007}
           }
\gridline{\rotatefig{270}{XRR050525A_monoflux_allband.eps}{0.24\textwidth}{(c) XRR050525A}
  \rotatefig{270}{XRR090423_monoflux_allband.eps}{0.24\textwidth}{(d) XRR090423}
          \rotatefig{270}{090102_monoflux_allband.eps}{0.24\textwidth}{(e) GRB090102}
          }
\caption{Multi-band light curves: (a) XRF050416A, (b) XRF081007, (c) XRR050525A, (d) XRR090423, and (e) GRB090102.  The data points are the observed light curves and solid lines are simulated light curves  with the boxfit tools (See  table \ref{tbl:boxfit_result}).\label{fig:boxfit_fit_result}}
\end{figure*}

 Table \ref{tbl:boxfit_result} summarizes the fitting results of the 5 samples that we used. We found the negative correlation between   $E_{{\rm iso}}$  (or $E_{{\rm peak}}^{{\rm src}}$)   and the opening angle of the jet, i.e., the sources with a smaller $E_{{\rm iso}}$ (or $E_{{\rm peak}}^{{\rm src}}$) have a larger opening angle.     The observing angles of  all the sample sources were 0$^{\circ}$.   These imply that these XRRs and XRFs are  the on-axis events.   Figure \ref{fig:boxfit_fit_result} shows the multi-band light-curves of  the samples. XRR050525A and XRR090423 showed the consistent fits to the X-ray and  optical data.  However, the  other three samples showed  unacceptable fits, especially  with the X-ray data.
 
 This result implies that an origin of X-ray afterglow emissions different from that of optical one.   The 
observed data of our samples are rather inconsistent with the predicted afterglow light-curve  on the basis of the external shock model. 

Although  our discussion has been based on the assumption that  both the X-ray and optical emissions  originate  in the external shock,  the possibility that 
the emissions  actually come from  some different  processes is not totally excluded.   Discussion about the possibility is out 
of scope of this paper.

\label{ss:boxfit_result}


\begin{deluxetable}{l|ccc}
\tablecaption{Samples  used for the boxfit simulations.\label{tbl:opt_sample}}
\tablehead{
  \colhead{Events} &  \colhead{Energy band or filters}  
}
\startdata
XRF050416A & X-ray\tablenotemark{a}, R\tablenotemark{b}, $K_{s}$\tablenotemark{c} \\
 XRF081007 & X-ray\tablenotemark{a}, r',H\tablenotemark{d} \\
 XRR050525A & X-ray\tablenotemark{a}, V,UVW1\tablenotemark{e} \\
 XRR090423 & X-ray\tablenotemark{a}, J,K,H\tablenotemark{f} \\
 GRB090102 & X-ray\tablenotemark{a}, r',z'\tablenotemark{g}\\
\enddata
\tablenotetext{a}{\citet{2009MNRAS...397..1177E}. We used the monochromatic flux at the 0.3 keV and 10 keV bands.}
\tablenotetext{b}{\citet{2007ApJ...661..982S}}
\tablenotetext{c}{\citet{2010ApJ...720..1513K}}
\tablenotetext{d}{\citet{2013ApJ...774..114J}}
\tablenotetext{e}{\citet{2006ApJ...637..901B}}
\tablenotetext{f}{\citet{2009Natur.461.1254T}}
\tablenotetext{g}{\citet{2010MNRAS...405..2372G}}
\end{deluxetable}

\begin{deluxetable*}{l c c c c c c c}
\tablecaption{Parameters derived by model fitting  with the boxfit tools. The fraction ($\epsilon_e$) of the downstream internal energy in the shock-accelerated electrons is fixed at $\epsilon_e=0.2$ for all the samples. The observing angle of the C-GRB (GRB090102) is fixed at $\theta_{obs}=0$.\label{tbl:boxfit_result}}
\tablehead{
  \colhead{Events} & \colhead{$\Delta \theta$\tablenotemark{h} [rad]} & \colhead{$\theta_{{\rm obs}}$\tablenotemark{i} [rad]} & \colhead{$E_{{\rm iso}}$ [$10^{52}$ erg]} & \colhead{$n$\tablenotemark{j} [cm$^{-3}$]}  & \colhead{$p$\tablenotemark{k}} & \colhead{$\epsilon_B$\tablenotemark{l} [10$^{-5}$]} & \colhead{$\chi^2/dof$} 
}
\startdata
 XRF050416A & 0.4949 & 0 & 0.05208 & 2.147 & 2.177 & 906.9 & 12.76 \\
 XRF081007 & 0.4136 & 0 & 0.0957 & 10.19 & 2.136 & 274.2 & 11.93 \\
 XRR050525A & 0.3817 & 0 & 1.608 & 34.38 & 2.100 & 14.91 & 19.61 \\
 XRR090423 & 0.05788 & 0 & 159.5 & 0.1331 & 2.697 & 0.8967 & 10.11 \\
 GRB090102 & 0.1909 & 0 (fixed) & 150.5 & 0.3802 & 2.164 & 0.0618 & 11.25 \\ 
\enddata

\tablenotetext{h}{Jet half-opening angle}
\tablenotetext{i}{Observing angle}
\tablenotetext{j}{Circum-burst number density}
\tablenotetext{k}{Synchrotron slope}
\tablenotetext{l}{The fraction of downstream internal energy in the shock-generated magnetic field}
\end{deluxetable*}


\section{CONCLUSIONS} \label{sec:conclusion}
\subsection{Conclusions about off-axis model}
We performed a systematic study of GRBs observed by \textsl{Swift} by investigating  the prompt and  afterglow emissions.  We  cataloged the long GRBs observed by the \textsl{Swift} between 2004 December and 2014 February,  classifying  them into three categories of XRFs, XRRs, and C-GRBs,  according to the classification method of \citet{2008ApJ...679..570S}.  We analyzed the spectra of these sources during the $t_{100}$ interval in the prompt emission and  derived $E^{{\rm obs}}_{{\rm peak}}$,  and also calculated $E^{{\rm src}}_{{\rm peak}}$ for those with known redshifts.  Analyzing X-ray afterglows of the GRB samples with well-constrained $E_{{\rm peak}}$,  we confirmed that $E_{{\rm peak}}^{{\rm src}}$ and $E_{{\rm iso}}$ are moderately correlated 
 with the X-ray luminosity and the temporal decay index.   Furthermore, we estimated total numbers of the XRFs, XRRs, and C-GRBs in the whole universe per year, to be $f_{{\rm C-GRB}}=570 \pm 36$,  $f_{{\rm XRR}}=3031 \pm 53$, and  $f_{{\rm XRF}}=968 \pm 45$  ($f_{{\rm XRF;E_{peak}>10keV}}=539 \pm 36$) [events yr$^{-1}$], respectively.   With these event rates, the canonical opening angle  $\Delta \theta$ and bulk Lorentz factor $\Gamma$ of the jets of the XRFs were estimated  to be $\Delta \theta \lesssim 0.3^\circ$ (corresponding to $\theta_{obs} \sim 0.6^{\circ}$), $t_{{\rm jet}}\lesssim30\,{\rm s}$ ($\lesssim40\,{\rm s}$) and $\Gamma\approx 1000$, respectively.   We thus conclude that the observer requires to be close to the jet on-axis for the XRFs.  This rejects one of the popular theoretical models,  the off-axis jet model, which proposes that the off-axis viewing angle of an observer to the jet plays an important role in the observed properties of the XRFs.  
\subsection{Suggestions from box-fit results}
We executed a model fit for the boxfit-simulated data  for the X-ray and optical afterglows in 2 XRFs, 2 XRRs, and 1 C-GRBs, respectively. This fitting results shows that the observing angles of all the sample sources were $0^{\circ}$, and this suggested that the XRRs and the XRFs are on-axis events. On the other hand, simulated multi-band light-curves in the samples showed unacceptable fits, especially with the X-ray data. The results implied that the external shock model alone could not explain the X-ray afterglows.

\section*{ }

This work is supported by MEXT KAKENHI Grant Numbers 17H06357 and 17H06362. The authors would like to thank R. Yamazaki for useful discussion. We also thank Y. Kawakubo, K. Senuma, and H. Ohtsuki for technical assistance with our analysis. 
Finally, we are grateful to the referees for useful comments.


\clearpage



\begin{thebibliography}{}

\bibitem[Amati et al.(2002)]{2002A&A...390..81A} Amati, L., et al. 2002, \aap, 390, 81
\bibitem[Amati et al.(2006)]{2006MNRAS...372..233A} Amati, L., 2006, \mnras, 372, 233
\bibitem[Band(1993)]{1993ApJ...413..281B} Band, D. L. 1993, \apj, 413, 281
\bibitem[Barraud et al.(2003)]{2003A&A...400.1021B} Barraud, C., et al. 2005, \aap, 400, 1021 
\bibitem[Barraud et al.(2005)]{2005A&A...440..809B} Barraud, C., Daigne, F., Mochkovitch, R., \& Atteia, J. L. 2005, \aap, 440, 809 
\bibitem[Barthelmy et al.(2005)]{2005SSRv...120..143B} Barthelmy, S. D., et al. 2005, \ssr, 120, 143
\bibitem[Berger et al.(2014)]{2014ARA&A...54..43B} Berger, E. 2014, \araa, 54, 43
\bibitem[Bloom et al.(1999)]{1999Natur.401..453B} Bloom, J. S. et al. 1999, \nat, 401, 453
\bibitem[Bloom et al.(2003)]{2003ApJ...588..945B} Bloom, J., Frail, D. A., Kulkarni, S. R. 2003, \apj, 588, 945
\bibitem[Blustin et al.(2006)]{2006ApJ...637..901B} Blustin, A. J. et al. 2006, \apj, 637, 901
\bibitem[Burrows et al.(2005a)]{2005SSRv...120..165B} Burrows, S. D., et al. 2005a, \ssr, 120, 165
\bibitem[Burrows et al.(2005b)]{10} Burrows, S. D., et al. 2005b, Science, 309, 1833B
\bibitem[Dermer et al.(1999)]{1999ApJ...513..656D} Dermer, C. D., et al. 1999, \apj, 513, 656
\bibitem[Donaghy(2006)]{2006ApJ...645..436D} Donaghy, T. Q. 2006, \apj, 645, 436
\bibitem[Evans et al.(2009)]{2009MNRAS...397..1177E} Evans, P. A., Beardmore, A. P., Page, K. L., et al. 2009, \mnras, 397, 1177
\bibitem[Frail et al.(2000)]{2000ApJ...537L..191F} Frail, D. A., Waxman, E., \& Kulkarni, S. R. 2000, \apj, 537, 191
\bibitem[Frail et al.(2001)]{2001ApJ...562L..55F} Frail, D. A., et al. 2001, \apj, 562, L55
\bibitem[Gehrels et al.(2004)]{2004ApJ...611..1005G} Gehrels, N., et al. 2004, \apj, 611, 1005
\bibitem[Gemdre et al.(2010)]{2010MNRAS...405..2372G} Gemdre, B., et al. 2010, \mnras, 405, 2372
\bibitem[Graff et al.(2016)]{2016ApJ...818...55G} Graff, P. B., Lien, A., et al. 2016, \apj, 818, 55
\bibitem[Gruber et al.(2014)]{2014ApJS..211...12G} Gruber, D., et al. 2014, \apjs, 211, 12
\bibitem[Heise et al.(2003)]{2003AIPC..662..229H} Heise, J. 2003, in AIP Conf. Proc. 662, Gamma-Ray Burst and Afterglow Astronomy 2001: A Workshop Celebrating the First Year of the HETE Mission, ed. G. R. Ricker \& R. K. Vanderspek (Melville, NY: AIP), 229
\bibitem[Heise et al.(2001)]{Heise01} Heise, J., Zand, J. I., Kippen, R. M., \& Woods, P. M. 2001, in Gamma-ray Bursts in the Afterglow Era, ed. E. Costa, F. Frontera, \& J. Hjorth (Berlin: Springer), 16
\bibitem[Jinet al.(2013)]{2013ApJ...774..114J} Jin, Zhi-Ping, et al. 2013, \apj, 774, 114
\bibitem[Kaneko et al.(2006)]{2006ApJS...116..29K} Kaneko, Y., et al. 2006, \apjs, 166, 29
\bibitem[Kann et al.(2010)]{2010ApJ...720..1513K} Kann, D. A., et al. 2010, \apj, 720, 1513
\bibitem[Kippen et al.(2003)]{2003AIPC..662..244K} Kippen, R. M., Woods, P. M., Heise, J., in’t Zand, J. J. M., Briggs, M. S., \& Preece, R. D. 2003, in AIP Conf. Proc. 662, Gamma-Ray Burst and Afterglow Astronomy 2001, ed. G. R. Ricker \& R. K. Vanderspek ( New York: AIP), 244
\bibitem[Kiziloglu et al.(2006)]{2006GCN..5618....1K} Kiziloglu, U., et al. 2006, GRB Coordinates Network, Circular Service, No. 5618, \#1
\bibitem[Kornilov et al.(2006)]{2006GCN..5901....1K} Kornilov, V., et al. 2006, GRB Coordinates Network, Circular Service, No. 5901, \#1
\bibitem[Lamb et al.(2005)]{2005ApJ...620..355L} Lamb, D. Q., et al. 2005, \apj, 620, 355
\bibitem[Lien \& Fields (2009)]{2009JCAP...1..47L} Lien, A., Fields, B. D. 2009, \jcap, 1, 47
\bibitem[Lien et al.(2014)]{2014ApJ...783..24L} Lien, A., Sakamoto, T., Gehrels, N., et al. 2014, \apj, 783, 24
\bibitem[Lien et al.(2016)]{2016ApJ...827..7L} Lien, A., Sakamoto T., Barthelmy, S. D., et al. 2016, \apj, 829, 7
\bibitem[Meegan et al.(1996)]{1996ApJ...106..65M} Meegan, C. A., et al. 1996, \apjs, 106, 65
\bibitem[Meegan et al.(2009)]{2009ApJ...702..791M} Meegan, C., Lichti, G., Bhat, P. N., et al. 2009, \apj, 702, 791
\bibitem[Paciesas et al.(1999)]{1999ApJS...122..465P} Paciesas, W. S., et al. 1999, \apjs, 122, 465
\bibitem[Racusin et al.(2008)]{2008Natur.455..183R} Racusin, J. L., Karpov, S. V., Sokolowski, M., et al. 2008, \nat, 455, 183
\bibitem[Racusin et al.(2016)]{2016ApJ...826..45R} Racusin, J. L., et al. 2016, \apj, 826, 45
\bibitem[P{\'e}langeon et al.(2008)]{2008A&A...491..157P} P{\'e}langeon, A., Atteia, J.-L., Nakagawa, Y.~E., et al. 2008, \aap, 491, 157
\bibitem[Rossi et al.(2002)]{2002MNRAS...332..945R} Rossi, E., et al. 2002, \mnras, 332, 945
\bibitem[Sakamoto et al.(2004)]{2004ApJ...602..875S} Sakamoto, T., et al. 2004, \apj, 602, 875
\bibitem[Sakamoto et al.(2005)]{2005ApJ...629..311S} Sakamoto, T., et al. 2005, \apj, 629, 311
\bibitem[Sakamoto et al.(2008)]{2008ApJ...679..570S} Sakamoto, T., et al. 2008, \apj, 679, 570
\bibitem[Santana et al.(2014)]{2014ApJ...785..29S} Santana, R., et al. 2014, \apj, 785, 29
\bibitem[Sari et al.(1999)]{1999ApJ...519L..17S} Sari, Re'em, Piran, Tsvi, \& Halpern, J. P. 1999, \apj, 519, 17
\bibitem[Schlegel et al.(1998)]{1998ApJ...500..525S} Schlegel, D. J., Finkbeiner, D. P., Davis, M. et al. 1998, \apj, 500, 525
\bibitem[Soderberg et al.(2006)]{2006Natur.442.1014S} Soderberg, A.~M., Kulkarni, S.~R., Nakar, E., et al. 2006, \nat, 442, 1014
\bibitem[Soderberg et al.(2007)]{2007ApJ...661..982S} Soderberg, A. M., et al. 2007, \apj, 661, 982
\bibitem[Spergel et al.(2007)]{2007ApJS...170..377S} Spergel, D. N., Bean, R., Dor, O., et al. 2007, \apjs, 170, 377
\bibitem[Tanvir et al.(2009)]{2009Natur.461.1254T} Tanvir, N. R., et al. 2009, \nat, 461, 1254
\bibitem[Urata et al.(2015)]{2015ApJ...806..222U} Urata, Y. Huang, K. Yamazaki, R. et al. 2015, \apj, 806, 222
\bibitem[van Eerten et al.(2012)]{2012ApJ...749...44V} van Eerten, H., van der Horst, A., \& MacFadyen, A. 2012, \apj, 749, 44
\bibitem[Wanderman et al.(2010)]{2010MNRAS...406..1944W} Wanderman, D., \& Piran, T. 2010, \mnras, 406, 1944
\bibitem[Woosley et al.(2006)]{2006ARA&A..44..507W} Woosley, S. E., \& Bloom, J. 2006, \araa, 44, 507
\bibitem[Yamazaki et al.(2002)]{2002ApJ...571L..31Y} Yamazaki, R., Ioka, K., \& Nakamura, T. 2002, \apj, 571, L31
\bibitem[Yamazaki et al.(2003)]{2003ApJ...593..941Y} Yamazaki, R., Ioka, K., \& Nakamura, T. 2003, \apj, 593, 941
\bibitem[Yamazaki et al.(2004)]{2004ApJ...607L..103Y} Yamazaki, R., Ioka, K., \& Nakamura, T. 2004, \apj, 607, L103
\bibitem[Yonetoku et al.(2004)]{2004ApJ...609..935Y} Yonetoku, D., Murakami, T., Nakamura, T., et al. 2004, \apj, 609, 935
\bibitem[Zang et al.(2006)]{2006ApJ...642..354Z} Zhang, B., et al. 2006 \apj, 642, 354

\end{thebibliography}
\end{document}